\documentclass[twocolumn]{aastex631}
\usepackage{amsmath}
\usepackage{tabularx}
\usepackage{bm}
\usepackage{amssymb}
\usepackage{graphicx}

\begin{document}

\title{Photometric analysis of asteroids in the Phocaea region}

\correspondingauthor{Xiaobing Wang}
\email{wangxb@ynao.ac.cn}

\author{Xiaoyun Xu}
\thanks{These authors contributed to the work equally and should be \\ regarded as co-first authors.}
\affiliation{Yunnan Observatories, Chinese Academy of Sciences, Kunming 650216, China}
\affiliation{University of Chinese Academy of Sciences, Beijing 100049, China}
\author{Xiaobing Wang}
\thanks{These authors contributed to the work equally and should be \\ regarded as co-first authors.}
\affiliation{Yunnan Observatories, Chinese Academy of Sciences, Kunming 650216, China}
\affiliation{University of Chinese Academy of Sciences, Beijing 100049, China}
\affiliation{Key Laboratory for the Structure and Evolution of Celestial Objects, Chinese Academy of Sciences, Kunming 650216, China}
\author{Karri Muinonen}
\affiliation{Department of Physics, P.O. box 64, FI-00014 University of Helsinki, Finland}
\author{Shenghong Gu}
\affiliation{Yunnan Observatories, Chinese Academy of Sciences, Kunming 650216, China}
\affiliation{University of Chinese Academy of Sciences, Beijing 100049, China}
\affiliation{Key Laboratory for the Structure and Evolution of Celestial Objects, Chinese Academy of Sciences, Kunming 650216, China}
\author{Antti Penttilä}
\affiliation{Department of Physics, P.O. box 64, FI-00014 University of Helsinki, Finland}
\author{Fukun Xu}
\affiliation{Yunnan Observatories, Chinese Academy of Sciences, Kunming 650216, China}
\affiliation{University of Chinese Academy of Sciences, Beijing 100049, China}
\affiliation{Key Laboratory for the Structure and Evolution of Celestial Objects, Chinese Academy of Sciences, Kunming 650216, China}
\author{Leilei Sun}
\affiliation{Yunnan Observatories, Chinese Academy of Sciences, Kunming 650216, China}
\affiliation{University of Chinese Academy of Sciences, Beijing 100049, China}
\affiliation{Key Laboratory for the Structure and Evolution of Celestial Objects, Chinese Academy of Sciences, Kunming 650216, China}
\author{Jing Huang}
\affiliation{Yunnan Observatories, Chinese Academy of Sciences, Kunming 650216, China}
\affiliation{University of Chinese Academy of Sciences, Beijing 100049, China}
\author{Pengfei Zhang}
\affiliation{Center for Lunar and Planetary Sciences, Institute of Geochemistry, Chinese Academy of Sciences, Guiyang, China.}
\author{Ao Wang}
\affiliation{Department of Physics, Yuxi Normal University, Yuxi 653100, China}

\begin{abstract}
The Phocaea asteroid family, one of the large ancient families located in the inner main belt, may be the sources of near-Earth asteroids (NEAs) due to the nearby 3:1 mean motion resonance with Jupiter, the $\nu 6$ secular resonance, and the Yarkovsky and YORP effects. 
Thus, understanding the influence of the Yarkovsky and YORP effects on the Phocaea family is one of the keys to figuring out the source of NEAs. 
However, the physical properties of most of the Phocaea family members are unknown at present. We perform a photometric analysis for 44 asteroids in the Phocaea region using photometric data obtained by ground-based and space-based telescopes (i.e., TESS and Gaia). 
Based on the derived physical properties, we find significant footprints of the Yarkovsky and YORP effects on the Phocaea family members. Selecting five asteroids nearby the inside boundary of the V-shape in the absolute-magnitude semimajor-axis ($H$, $a$) space, we estimate their densities considering their migration in semimajor-axis arises from the Yarkovsky effect. The bulk density of (852)~Wladilena (3.54~$g/cm^{3}$) suggests a link to the H chondrite meteorites. Incorporating the grain density of the H chondrites, we estimate the macroporosities of the asteroids (290)~Bruna, (1164)~Kobolda, and (587)~Hypsipyle, respectively $41\%, 47\%$, and $65\%$, implying rubble pile structures. 
Considering the H chondrites link to asteroid (25)~Phocaea, we suggest the parent body of the Phocaea family has been composed of H chondrite like material and the Phocaea family may be one of the sources of H chondrite meteorites.

\end{abstract}

\section{Introduction}
Asteroids with similar orbits, i.e., proper semimajor axes, eccentricities, and inclinations, can constitute a  dynamical family. Such families, also called asteroid families, are thought to have their origins in catastrophic collisions of larger parent bodies. Family members, fragments of such collision events, have been subject to both gravitational and non-gravitational effects, such as the Yarkovsky and YORP effects. Of course, they have also experienced space weathering, solar heating, and maybe subsequent collisions. The physical properties and orbits of these asteroids have been shaped by the above mentioned effects \citep{Masiero2015}. To an extent, asteroid families are valuable laboratories for humans to study planetary-scale impact physics, which could provide us with important insight into the task of planetary defense, as well as help us to figure out the origin and evolution of asteroids \citep{Novakovic2022}.

We are interested in the Phocaea dynamical family for several reasons. Firstly, it is one of the largest ancient asteroid families formed by a collisional outbreak \cite{Carvano2001}. Secondly, its special location in the solar system, near three resonances (the 7:2 and 3:1 mean motion resonances with Jupiter and the $\upsilon_6$ secular resonance), suggests that it can be a source of near-Earth asteroids \citep[NEAs,][]{Bottke2000, GRANVIK2018,Marsset2022}. Thirdly, the Phocaea dynamical region will be covered by the survey of the Chinese Space Station Telescope (CSST), by which multiple band photometric data for asteroids of apparent magnitude brighter than 26 mag and spectroscopic data of wavelength range from 0.255 $\mu m$ to 1.0 $\mu m$ for asteroids brighter than 23 mag can be obtained \citep{zhan2021}. 
Asteroid (25)~Phocaea is the largest member of the Phocaea family, also one of the largest known S-type asteroids the inner main belt of asteroids.
At present, there are 5796 members of the Phocaea family, according to information on the Asteroid Families Portal\footnote{\url{http://asteroids.matf.bg.ac.rs/fam/}}. Actually, discussing Phocaea region is preferable (an orbital region of semimajor axes ranging from 2.2 to 2.5~au, eccentricities between 0.1 and 0.3, and inclinations from 17 to 27 degrees), because the Tamara family, composed of dark asteroids, is also located in this region.
About 87\% of the Phocaea family members are fainter than 15~mag. Due to observational constraints, members with known physical properties are rare.
\cite{Carvano2001} firstly determined the spectral types of 31 asteroids in the Phocaea region, and found that 24 members are S-type, 3 are C-type, 3 are X-type, and 1 is D-type.
\cite{Carruba2009} analyzed the spectra of 50 asteroids in the Phocaea region: 36 are identified as S-type and 4 as C-type, with 10 belonging to other types.
Most of the information on the members of the Phocaea region comes from the observations of the NEOWISE\footnote{\url{https://sbn.psi.edu/pds/resource/neowisediam.html}} mission, by which the diameters and albedos of 1307 asteroids in the Phocaea region are determined. Among 1307 asteroids,  85\% asteroids have diameters smaller than 5~km and 62\% have an S-type geometric albedo (ranging from 0.15 to 0.4). The percentage of asteroids with S-type albedo is consistent with previous studies \citep{Carvano2001,Carruba2009} showing that the Phocaea region is predominantly composed of S-type asteroids.
Adding albedo information of asteroids, \cite{Novakovic2017} identified a new asteroid family consisting of 226 low-albedo asteroids in the Phocaea region, which imply a more complex family structure in the region. 
For the age of the Phocaea family, \cite{Carruba2009} provided an upper limit of 2200 Myr using the method of \cite{VOKROUHLICKY2006}. \cite{Milani2017} determined its age as $1187\pm319$~Myr by fitting the inner side of the so-called V-shape in $(1/D,a)$ plane ($D$ and $a$ are the diameter and semimajor axis).

As for the spin and shape of the members of Phocaea family, \cite{Carruba2009} determined rotation periods for 18 asteroids. With  8 out of the 18 asteroids being slow rotators, the excess of slow rotators was considered to be a result of the YORP effect. \cite{Hanus2013} have carried out photometric inversion for 14 members of the Phocaea family, finding significantly more retrograde  than prograde rotators. Obviously, to understand the origin and evolution of the asteroids in the Phocaea region, it is necessary to derive basic physical properties such as size, shape, spin, composition, and even density for a larger number of asteroids in the region. 

Nowadays, time-domain space-based and ground-based surveys are running or being prepared, providing us opportunities to extend the physical properties of asteroids and to uncover puzzles of the origin and evolution of individual asteroids, asteroid families, and even the entire asteroid main belt. For example, Gaia DR3 provided calibrated G-band sparse photometric data for more than 150,000 asteroids with multiple observed geometry and spectroscopic data for 60,518 asteroids \citep{Tanga2023}. Based on Gaia DR3 photometric data and ground-based photometric data, \cite{Cellino2024} determined the rotation periods, pole orientations, shapes, and photometric slopes for over 22,000 asteroids. \cite{Martikainen2021} did photometric inversion for 491 asteroids with the Gaia DR2 data and ground-based data from the Database of Asteroid Models from Inversion Techniques (DAMIT). The Transiting Exoplanet Survey Satellite (TESS) mission, a time-domain survey dedicated to searching for new transit exoplanets, provides dense lightcurves of 9912 asteroids \citep{pal2020}. The Yunnan-Hong Kong wide field photometric (YNHK) survey, a ground-based time-domain survey also dedicated for new transiting exoplanets, has offered dense lightcurves for 546 asteroids \citep{Gu2022,xu2023}. 

By performing lightcurve inversion, in the present work, we determine the physical parameters of 44 asteroids in the Phocaea region with photometric data from Gaia DR3, TESS data release, YNHK survey, and the DAMIT database. It is worth mentioning that the absolute magnitude and photometric slope of the 44 asteroids are determined because the calibrated magnitudes (e.g., Gaia DR3, TESS data release, and YNHK data) of selected asteroids are involved. Based on the derived spin parameters of 44 asteroids, we may have the opportunity to investigate the influence of the YORP and Yarkovsky effects on the Phocaea family.  

Therefore, this paper is organized as follows. Section 2 is the information on the photometric observations and data reduction for 44 selected asteroids in the Phocaea region. The photometric inversion methods used in this work are shown in Section 3. The photometric inversion results of 44 asteroids and related discussions are presented in Section 4. Finally, a summary is given in the last section. 

\section{Photometric Data}
Photometry is the most economical and efficient way to determine the physical properties of asteroids. The value of brightness of an asteroid at a certain time and the shape of its lightcurve (brightnesses at different time) are related to the asteroid's physical properties (size, shape, spin parameters, and surface scattering parameters) and observational geometries. That means from both dense lightcurves and sparse calibrated brightnesses of asteroids we can extract their related physical parameters, just like \cite{Martikainen2021}'s work shown. In this work, we performed lightcurve inversion for 44 selected asteroids using their dense lightcurves and sparse photometric data, some of which are obtained by ground-based and space-based surveys.
The sparse photometric data of 44 involved asteroids used in this work comes from Gaia DR3, which are calibrated G-band magnitudes covering phase angles ranging from $10^{\circ}$ to $40^{\circ}$ for main-belt asteroids \citep{Tanga2023}.
The dense photometric data of the 44 involved asteroids come from data release of the TESS survey, the YNHK survey, the DAMIT database \citep{Durech2010}, and the Asteroid Lightcurve Data Exchange Format (ALCDEF) Database of the Minor Planet Center \citep{Warner2009}.

YNHK survey employed a Centurion 18-inch telescope attached a clear filter at the Lijiang station of Yunnan Observatories, is the time-domain survey dedicated to new exoplanets searching. YNHK survey has been run more than 7 years by our team \citep{Gu2022}. To search new transit exoplanets from the YNHK survey, \cite{Gu2022} built a data reduction pipeline which contains four parts: basic image processing, astrometry calibration and cross-match with a certain catalog, photometric measurement, and read noise correction. In order to extract dense lightcurves of asteroids from the YNHK survey, we  developed a special sub-pipeline using machine learning techniques \citep{xu2023}. With our special sub-pipeline, asteroids are identified with their celestial coordinates and velocities. Then, the data reduction procedure of asteroids in the YNHK survey is the same as that of stellar objects when the coordinates of asteroids identified by our special sub-pipeline are added into the input catalog which is customized according to the specific scientific goals. The detailed information can refer to papers \citep{Gu2022,xu2023}. A recently update for our pipeline is that the Gaia DR3 catalog (Gaia Data Release 3) is applied in the astrometric and photometric  calibrations. 

The TESS survey provides calibrated dense lightcurves of 9912 asteroids at a cadence of 30 minutes \citep{pal2020}. Compared to ground-based observations, the TESS survey has the merit of continuous 24-hour observation, which is important for physical studies of asteroids with a long rotational period. Before the photometric analysis, the calibrated TESS magnitudes $m_{T}$ are converted into the Gaia G-band magnitudes $m_{G}$ with the aid of the transformation relationships provided by \cite{Stassun_2018,GaiaEDR3_2021}:
\begin{equation}
    m_{G}=m_{T}+0.2991-0.0057*(g-i)+1.298*(r-i),
    \label{t2g}
\end{equation}
where the color indices $(g-i)$ and $(r-i)$ of the asteroids come from the database of the Sloan Digital Sky Survey \citep[SDSS,][]{ivezic2020sdss}.

The dense lightcurves of our targets are from the ALCDEF and DAMIT databases. The ALCDEF database contains photometric data for more than 24,000 asteroids. The lightcurves of 44 selected asteroids from the ALCDEF database have a quality code $U \geq 2$. 

For all photometric data included, the light travel times of asteroids are corrected for and the apparent magnitudes of the asteroids are converted into reduced magnitudes, that is, the values at 1~au distances from the Sun and the observers. 

In the photometric analysis for 44 asteroids in the Phocaea region, 453 dense lightcurves and 1105 sparse photometric data from the aforedescribed datasets are included. In detail, 330 dense relative lightcurves of 37 asteroids, observed by ground-based telescopes, were downloaded from the ALCDEF and DAMIT databases.
These data have time intervals ranging from one minute to several minutes. The 121 lightcurves of 9 asteroids from the TESS survey and 2 dense lightcurves of the asteroid (1626) Sadeya come from the YHNK survey. Sparse photometric data of asteroids mainly come from Gaia DR3, there are at least 12 data points at more than three different phase angles for each asteroid.

\section{Photometric inversion}
To invert the physical properties of the Phocaea family members from the photometric data, we perform Bayesian lightcurve inversion developed by \cite{Muinonen2020, Muinonen2022}. The disk-integrated brightness of an asteroid at a certain observational geometry is derived by the sum of radiation reflected from the illuminated and visible surface area of the asteroid.  Thus, the brightness model of the asteroid in the Bayesian lightcurve inversion method involves a convex shape represented with truncated spherical harmonics, the Lommel-Seeliger scattering model, and the $H,\!G_1,\!G_2$ phase function. Setting a triangulation for the unit sphere, the observed brightness of an asteroid at anytime can be calculated as following formal sum,
\begin{equation}
\begin{aligned}
   L(E,E_0,\alpha)=\Sigma S(\mu_i,\mu_{0,i})G(\vartheta_i,\psi_i)\sigma_i,\\
  S(\mu,\mu_{0})=2p\frac{\Phi(\alpha)}{\Phi_{LS}(\alpha)}\frac{1}{\mu+\mu_{0}},\\
   \mu=E\cdot n, \; \mu_0=E_0\cdot n, 
\end{aligned}
\end{equation}
where $E, E_0$ represent the directions of the observer and source and $G(\vartheta_i,\psi_i)\sigma_i$ denotes the size of the facet with the unit normal vector $n(\vartheta_i,\psi_i)$. The Gaussian surface density function $G(\vartheta,\psi)$ is approximated with truncated spherical harmonics in the practical computation. $\Phi_{LS}(\alpha)$ denotes the Lommel-Seeliger phase function (see Eq.~(6) in \cite{Muinonen2020}) and $\Phi(\alpha)$ is a linear phase function (see Eq.~(10) in \cite{Muinonen2022}).

Considering that the asteroid spins about its axis of maximum inertia, the unknown parameters in the brightness of the asteroid can be $\textbf{\textit{P}}=(T,\lambda_p,\beta_p,\varphi_0,s_{00},\ldots,s_{l_{\textit{max}}l_{\textit{max}}},m_{20},\beta_{s})$. In detail, $T,\lambda_p,\beta_p,\varphi_0$ are the spin parameters of the asteroid.
If one sets $l_{\textit{min}} = 0$ and $l_{\textit{max}} = 6$, there will be 49 spherical harmonics coefficients or, say, shape parameters $s_{00},\ldots,s_{l_{\textit{max}}l_{\textit{max}}}$. The phase function parameter $\beta_{s}$ and $m_{20}$ are involved because of application to absolute Gaia data.
The determination procedure for the unknown parameters with the Bayesian inversion actually contains the least-squares fitting and the MCMC analysis parts. The chi-square sum is computed with a weighting scheme determined by the quality of data and the duration and time resolution of the observations,
\begin{equation}
\begin{aligned}
    \chi^{2}(\textbf{\textit{P}}) \approx & \sum_{i=1}^{4} \sum_{k=1}^{K_{i}} \frac{(2.5 \, \mathrm{\log_{10}~e})^2}{\sigma_{\epsilon,ik}^2} \\
    & \times \sum_{j=1}^{N_{ik}} \left[\frac{L_{\textit{obs},ikj} - L_{ikj}(\textbf{\textit{P}}) \, 10^{0.4 \Delta M_{ik0}(\textbf{\textit{P}})}}{L_{\textit{obs},ikj}}\right]^2, \\
    \sigma_{\epsilon,ik}=& \sqrt{\frac{N_{ik}}{N_{ik,\textit{eff}}}} \max(\sigma_{0,ik},\sigma_{pr,ik}), \quad
\end{aligned}
\end{equation}
where $N_{ik}$ and ${\sigma_{\epsilon,ik}}$ are the number of observations and the weight of the $ik$-th lightcurve, $L_{\textit{obs},ikj} $ and $L_{ikj}(\textbf{\textit{P}})$ are the observed and computed brightnesses, $\Delta M_{ik0}(\textbf{\textit{P}})$ is the averaged difference between observed and computed magnitudes. The index $i$ in the above formula represents the characteristic of each lightcurve, $i=1,2,3,4 $ refer to dense relative, sparse relative, dense absolute, and sparse absolute lightcurve respectively. The dense and sparse lightcurves are identified according to the sampling rates of the lightcurves in time. $N_{ik,\textit{eff}}$ is the so-called efficient number of observations in the lightcurve. For relative lightcurve. $N_{ik,\textit{eff}}$ is estimated by comparing the mean sampling time interval of the lightcurve to that of the most sparse relative lightcurve. For absolute lightcurves, $N_{ik,\textit{eff}}$ is set to unity.
The initial values of $\sigma_{0,ik}$ are set as the standard deviation of the spline fitting to the raw data. Then, they are related to the root mean square (RMS) of the O-C residuals between the observed and computed brightnesses. $\sigma_{pr,ik}$ is a prior threshold of uncertainty, set to 0.005~mag for dense lightcurves and 0.02~mag for the Gaia data.
  
Firstly, we apply the Levenberg–Marquardt algorithm to find the least-squares solution for the unknown parameters. During the least-squares fitting procedure, the involved parameters $\textbf{\textit{P}}$ and ${\sigma_{\epsilon,ik}}$ are refined iteratively until they converge. 

Secondly, the virtual observation MCMC analysis for the photometric data of 44 asteroids is done using the Metropolis-Hastings algorithm. During the virtual observation MCMC procedure, a set of virtual observations $M_v$ is generated from the original observations $M_{obs}$ by adding additional Gaussian random quantities of given uncertainties $\nu$ and a covariance matrix $\Lambda$. A set of virtual least-squares solutions of unknown parameters are derived from the virtual observation sets. Then, a random-walk  MCMC sampling is performed from the symmetric proposal probability density function (p.d.f.) composed of the virtual least-squares solution of parameters. More detailed mathematical description can be found in \cite{Muinonen2020}. Based on a posteriori p.d.f.s  of involved parameters, the best parameter values and their uncertainties are obtained.

Additionally, we estimated the G-band absolute magnitudes $H_G$, and phase function parameters $G_1,\!G_2$ of 44 asteroids with the method  of \cite{Martikainen2021}. In practice, these parameters are derived by fitting the phase curves of 44 asteroids at reference geometry (equivalent an equatorial illumination and observation) with the help of convex shape solutions derived in the MCMC procedure. Because only the Gaia data are absolute photometry in the analysis, the slopes $\beta_{\textit{ref}}$ and $H_{G}$ of the mean-magnitude reference phase curve are estimated \citep{Muinonen2022}.

In the least-squares fitting,  the initial rotation periods of 44 asteroids are retrieved from the Asteroid Lightcurve Database \citep[LCDB\footnote{\url{https://minplanobs.org/mpinfo/php/lcdb.php}},][]{WARNER2009134} or DAMIT database. The initial rotational phase $\varphi_0$ is typically set to zero at a given epoch $t_0$ where generally is the brightness minimum. The initial pole orientations of 42 members were obtained from the DAMIT database. The asteroids (8356)~Wadhwa and (23552)~1994~NB do not have prior estimates of their pole orientation and shape. For these two asteroids, a systematic scan with 10-degree mesh over the unit sphere is performed to find their possible pole solutions with the least-squares method. 
Once the initial values of the spin parameters are determined, the optimization algorithm is applied to iteratively refine the shape and phase function parameters. 

%

%
\section{Results and Discussion}
\begin{figure*}[ht]
    \centering
    \includegraphics[width=0.49\textwidth]{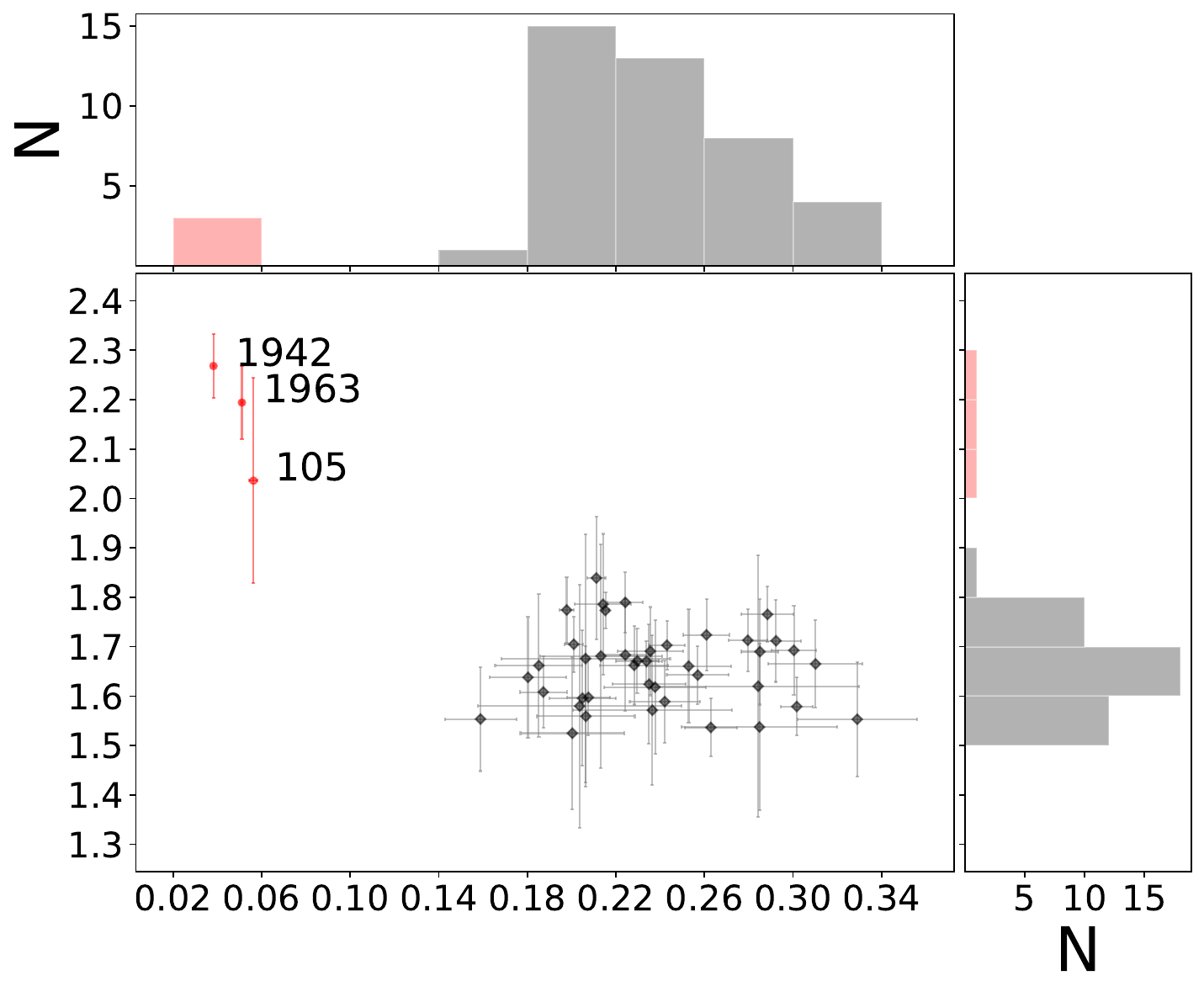}
    \includegraphics[width=0.49\textwidth]{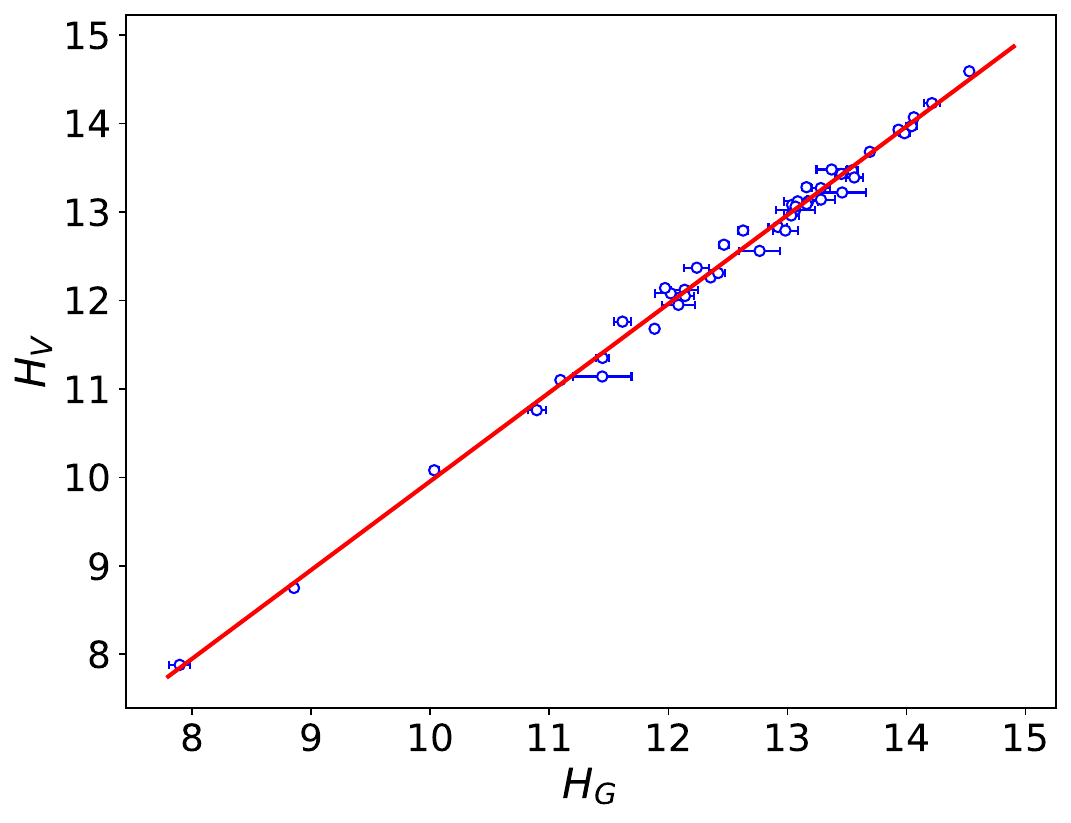}
    \caption{Left: the distributions of the G-band geometric albedo $p_{_G}$ and photometric slope $\beta_{\textit{ref}}$ for 44 asteroids.  Right: the distribution of $H_G$ vs. $H_V$ for 44 asteroids. The red line represents the linear relationship $H_V=H_G$.}
    \label{HG}
\end{figure*}
\subsection{The spin parameters and shapes}
Applying the Bayesian lightcurve inversion method, we analyze the photometric data of 44 asteroids in the Phocaea region, and obtain their convex shape, spin parameters, and phase function parameter.  
The best values for the spin parameters are presented in Table \ref{Information on spin parameters}, the photometric slope $\beta_{\textit{ref}}$ with their uncertainties are presented in Table \ref{Information on physical parameters}.
Using the Minkowski problem solver, the convex shape of the best-fit solution has been reconstructed (see Figure \ref{shape}). 

Among involved asteroids, we obtained the unique pole solution for 40 asteroids in this photometric inversion, for 4 other asteroids, (587) Hypsipyle, (6510) Tarry, (23552) 1994 NB and (29729) 1999 BY1, we could not rule out their mirror pole solutions (a pair of pole solutions at similar latitudes but with longitudes differing by $180^\circ$). The ambiguity may arise from the limitation of the observational geometry of the photometric data used.  In practice, we generally tested their possible mirror pole for all involved asteroids, then compared the fitting situations for the pair of mirrored pole solutions. Finally, we selected the admissible pole solution for each asteroid by considering fitting situation of modeled brightnesses to observed ones, especially to the Gaia data. 
Statistically speaking, we determined the basic physical parameters of 2 asteroids for the first time, and derived the more precise parameters for other 42 asteroids. For examples, the uncertainties of rotation period for 44 asteroids range from  $10^{-7}$ to $10^{-5}$ hours,  the uncertainties of pole orientations in both ecliptic longitude and latitude are less 5 degrees. 

As examples, the lightcurve inversion results are shown for  four asteroids: (1626) Sadeya, (6560) Pravdo, (8356) Wadhwa, and (23552) 1994 NB.

\textbf{(1626) Sadeya}

Several groups have carried out  physical studies for (1626) Sadeya, deriving different pole solutions.  \cite{Durech2020AA...643A..59D} derived a pole solution of $(250^{\circ}, -37^{\circ})$, with a rotation period of 3.42015 hr. \cite{Stephens2021MPBu...48...56S}'s study suggested a pole solution of $(152^{\circ}, -9^{\circ})$ of a rotation period of 3.421367 hr.  In this analysis,  14 dense lightcurves of Sadeya and 16 calibrated Gaia data are involved. These dense lightcurves are obtained by ground-based telescopes at five apparitions: 2007, 2009, 2014, 2018, and 2020, with the phase angle of  the Gaia data ranging from $17^{\circ}$ to $25^{\circ}$. We obtained the best pole solution of $(251.7^{\circ}\pm1.5^{\circ}, -43.2^{\circ}\pm1.4^{\circ})$ with a period of $3.4201645\pm0.0000005$ hr. Figure \ref{lclc} (a) shows  example photometric data (blue circles) for Sadeya (left: a dense lightcurve, right: Gaia data) together with the model points (red symbols) computed by using the chosen admissible pole solution represented. The observed-minus-computed (O-C) residuals are shown in the bottom panel of Figure \ref{lclc} (a). For the chosen pole solution, the $\chi^{2}$ value is 28.7 and the rms-value of Gaia data is 0.0024 mag. For the mirror pole solution $(65.5^{\circ}\pm1.1^{\circ}, 19.8^{\circ}\pm1.7^{\circ})$, $\chi^{2}$-value is 30.2 and the rms-value of Gaia data is higher at 0.0127 mag for the Gaia data. Thus, we consider the former pole as the preferred pole solution. 

\textbf{(6560) Pravdo }

Physical studies for (6560) Pravdo have been restricted because of the limited availability of photometric data. Using sparse data from the ATLAS survey, \cite{Durech2020AA...643A..59D} roughly estimated the ecliptic latitude of the pole as $56^\circ$ with an uncertainty of $18^\circ$. Recently, \cite{Durech2023} determined its shape and pole parameters using Gaia data. The new derived pole orientation is $(174^{\circ}, 48^{\circ})$ with a rotation period of 19.1981 hours. Using 17 dense lightcurves obtained by TESS and 33 Gaia data points, we analyze the physical parameters of (6560) Pravdo. The dense lightcurves from TESS of a 30-min cadence are obtained consequently within 14 days. The 33 Gaia points cover phase angles from $15^{\circ}$ to $29^{\circ}$. We obtained the best values for spin parameters of ($174.2^{\circ}\pm1.0^{\circ}, 51.4^{\circ}\pm2.3^{\circ}$) with a rotation period of $19.1980260\pm0.0000201$ hr, close to the results reported by \cite{Durech2023}.  The $\chi^{2}$-value of the best pole solution is 47.1, and the rms-value of the Gaia data is 0.0071 mag. In comparison, the mirror pole solution of ($348.8^{\circ}, 88.7^{\circ}$) gives the rms-value of 0.0257 mag for Gaia data, which is significantly larger than that for the best pole solution, so it is rejected. Figure \ref{lclc} (b) presents example photometric data with computed values for (6560) Pravdo. The shape of the lightcurves for Pravdo shows a regular double-peaked sinusoidal structure with a high amplitude close to 0.75 mag. Consequently, an elongated shape suggesting binary characteristics is derived (see Figure \ref{shape}).    

\textbf{(8356) Wadhwa}

The physical parameters of (8356) Wadhwa are here derived for the first time by us. In the present  photometric analysis, 11 dense lightcurves and 28 Gaia data points of (8356) Wadhwa are used. The dense lightcurves, observed at three different apparitions, are collected from \cite{Buchheim2009MPBu...36...84B} and the ALCDEF database. We obtained the best-fit pole solution of $(63.8^{\circ}\pm1.7^{\circ}, 74.4^{\circ}\pm1.3^{\circ})$ with a period of $ 3.0432333\pm0.0000005$ hr. Figure \ref{lclc} (c) shows example lightcurves for (8356) Wadhwa. The solution has a $\chi^{2}$-value of 37.7 and an rms-value of 0.0107 mag for Gaia data.
The mirror pole of ($251.5^{\circ}, 89.1^{\circ}$) is rejected because of a larger rms-value (0.0326 mag) for the Gaia data.

\textbf{(23552) 1994 NB}

Our work marks the first time that the physical parameters of (23552) 1994 NB are analyzed.
For 1994 NB, we used 28 Gaia data points and 6 dense lightcurves observed in 2012 and 2023, published by \cite{Gonz2023MPBu...50..258L} and \cite{Brian2023MPBu...50...74S}. In our analysis, we derived a pair of pole solutions with similar $\chi^2$-values: $(321.0 ^{\circ}\pm4.6^{\circ}, -73.2^{\circ}\pm0.9^{\circ})$ and $(164.0 ^{\circ}\pm2.2^{\circ}, -83.1^{\circ}\pm0.7^{\circ})$. For the poles, similar rotation periods of about $3.628687$ hr are derived. The corresponding $\chi^{2}$-values are 8.7 and 9.6 and the rms-values of the Gaia data are 0.0093 and 0.0139 mag, respectively. Accordingly, we could not indicate a preferred pole solution for the 1994 NB. Figure \ref{lclc} (d) shows example modeled lightcurves of 1994 NB for both pole solutions, indicating the solutions ($321^{\circ}$, $-73^{\circ}$) and ($164^{\circ}$, $-83^{\circ}$) with red and black symbols, respectively.

For all targets of this work, the input photometric data, derived model parameters and figures of lightcurves fitting are available via China-VO PaperData repository\footnote{\url{https://nadc.china-vo.org/res/r101597/}}. The spin parameters of 44 asteroids derived in this work are shown in Table~\ref{Information on spin parameters}, along with the corresponding values from literatures. Among our samples, 21 were previously reported of two or more possible pole orientations. After our analysis, the unique pole for 19 of them are derived. The rotation periods of 41 asteroids are consistent with previously reported values. For (587) Hypsipyle, we confirmed its rotation period as 2.8894654 hr which is differents from the 13.6816 hr in \citet{Hanus2016}, but agreement with the spin period given by \citet{Durech2020AA...643A..59D}.
Among the 44 asteroids, the pole solutions of 15 asteroids are close to previous results. While some asteroids, for examples, (1942) Jablunka, (7055) Fabiopagan, and (56086) 1999 AA21, have different pole orientations from previous results, particularly in pole ecliptic longitude.


\subsection{Phase function analysis}
\begin{figure*}[ht]
    \centering
    \includegraphics[width=0.49\textwidth]{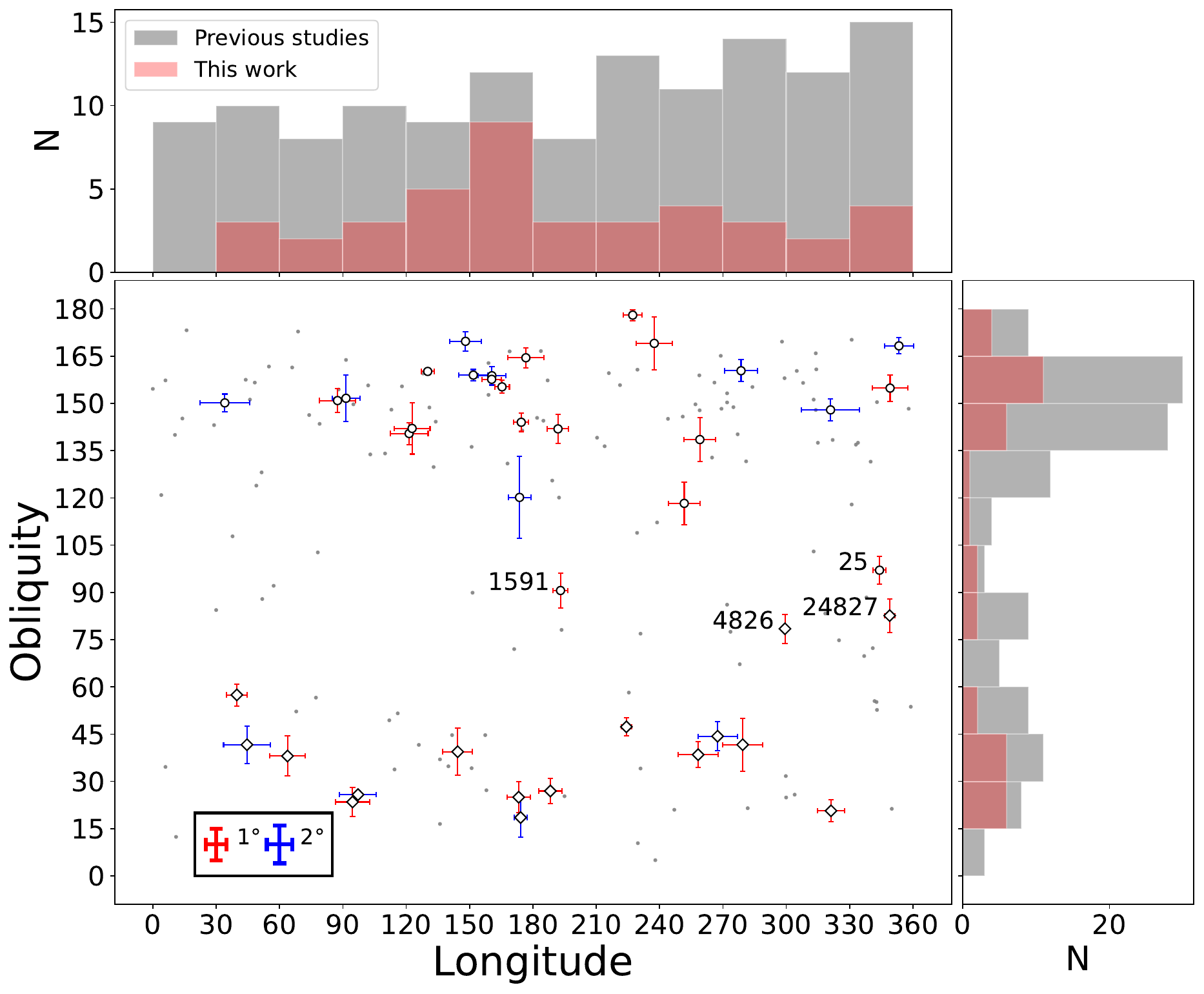}
    \includegraphics[width=0.49\textwidth]{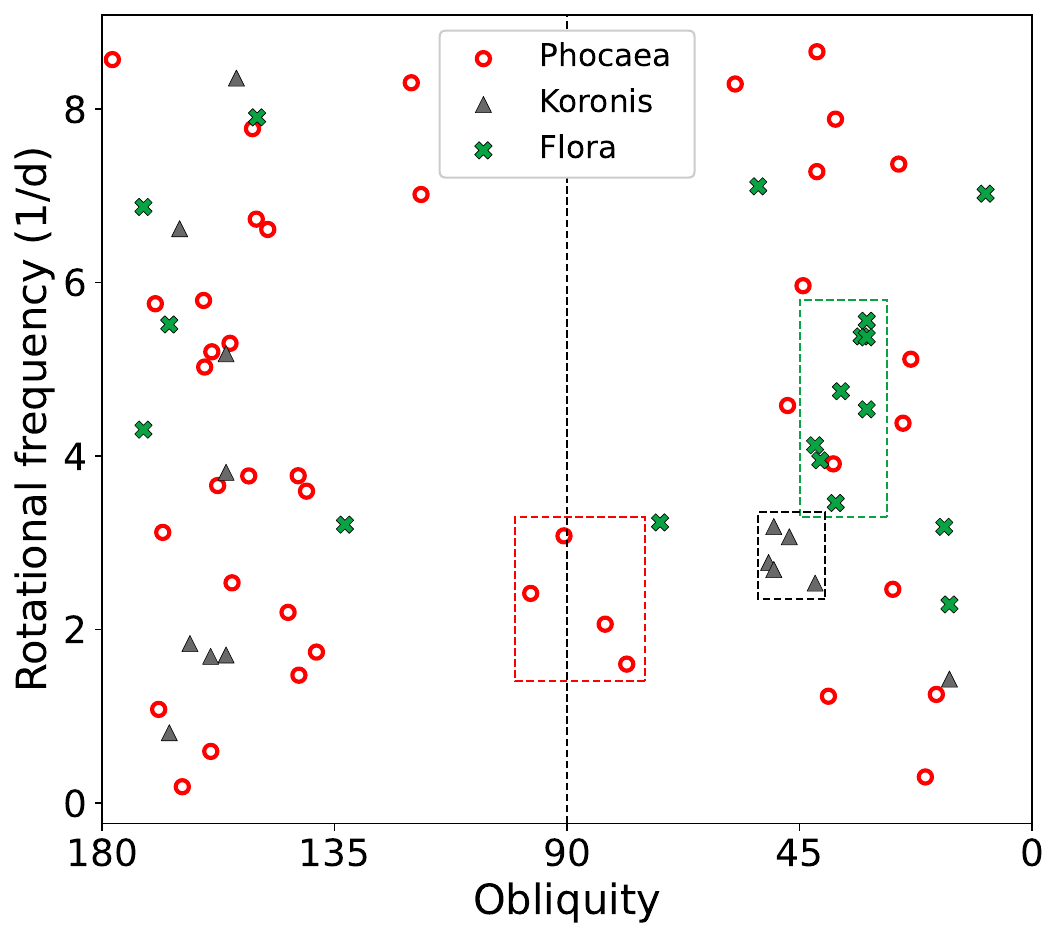}
    \caption{Left: the distributions of pole longitudes and obliquities for 172 asteroids (41 derived from our work and 131 from previous studies). right: the distribution of obliquities vs. spin rates for the 41 asteroids. 
        }
    \label{Dist-pole}
\end{figure*}
During photometric inversion, we also derived the absolute magnitudes $H_G$ and photometric slopes $\beta_{\textit{ref}}$ of 44 asteroids in the Gaia G band; these are listed in Table \ref{Information on physical parameters}. From the left panel of Figure \ref{HG}, the values of $\beta_{\textit{ref}}$ for 41 asteroids are between 1.5 and 1.85, which correspond to S-complex class according to the results by \cite{Martikainen2021} and \cite{Muinonen2022}. The three remaining asteroids (105) Artemis, (1942) Jablunka, and (1963) Bezovec have large values of $\beta_{\textit{ref}}>2.0$, which implies C-complex class. These results support the hypothesis that S-complex asteroids predominate in the Phocaea region \citep{Carvano2001}. 
As for the three C-complex asteroids, asteroid (105) Artemis and (1963) Bezovec are classified into C-complex by spectral data and near-infrared color \citep{BUS2002146, Tholen1984PhDT.........3T, Popescu2018}.  
The asteroid (1942) Jablunka, with the color indices (g-r), (g-i), and (i-z) of 0.2988, 0.4040, and 0.0304 \citep{Sergeyev2022}, is also consistent with C-complex classification \citep{DEMEO2013}. Actually, \cite{Carruba2009} declare (105) Artemis as being an interloper of the Phocaea family. In our view, the asteroids (1942) Jablunka and (1963) Bezovec are not members of the Phocaea family either. Asteroid (1942) Jablunka of diameter of 16.7 km has been confirmed as the third largest member of the Tamara family \citep{Novakovic2017}. Considering the large sizes of the asteroids (105) Artemis and (1963) Bezovec (94 km and 35 km, respectively) and the size of the parent of the Tamara family (106.1 km, estimated by \cite{Novakovic2017}), they cannot belong to the Tamara family.

The relation between the $H_G$ magnitudes and the V-band absolute magnitudes $H_V$ given by the Jet Propulsion Laboratory (JPL) service for the 44 asteroids is shown in the right panel of Figure \ref{HG}. The high correlation coefficient of 0.9964 reflects the fact that the $H_G$ values are consistent with the $H_V$ system.

Using the derived G-band absolute magnitudes ($H_G$) and asteroid diameter ($D$) from the NEOWISE survey, we estimated their geometric albedos ($p_{_G}$) in the G band of the 44 asteroids by Eq.(\ref{calp}), because the Gaia G band covers a larger portion of the incident solar spectrum \citep{pentikainen2024asteroid}.
\begin{equation}
    \log{p_{_G}} = 6.247-0.4H_G-2\log{D}
    \label{calp}
\end{equation}
The estimated geometric albedos $p_{_G}$ are listed in the last column of Table \ref{Information on physical parameters}. The uncertainties of the geometric albedos  are estimated based on the errors of $H_G$. For the 41 asteroids with slope parameters smaller than 1.85, as shown in the left panel of Figure \ref{HG}, the geometric albedos $p_{_G}$ are between 0.158 and 0.328 with a mean value of 0.238, which is consistent with typical geometric albedos of S-complex asteroids. The geometric albedos of asteroid (105) Artemis, (1942) Jablunka, and (1963) Bezovec are 0.056, 0.038, and 0.051, respectively, which is consistent with the geometric albedos of C-complex asteroids. Thus, we believe that the aforedescribed 41 asteroids are members of the Phocaea family.

\subsection{Footprints of the Yarkovsky and YORP effects}

To investigate the footprints of the Yarkovsky and YORP effects on the Phocaea family,  only the derived physical parameters of the 41 S-complex asteroids are used. Firstly, we analyze the pole distribution of the asteroids, as shown in the left panel of Figure \ref{Dist-pole}, we find that these poles tend to concentrate onto two distinct clusters in obliquity (angle between the rotational pole orientation of an asteroid and the normal vector of its orbital plane). 
The prograde cluster contains asteroids with obliquities ranging from $0^\circ$ to $45^\circ$, and the retrograde cluster consists of asteroids with obliquities between $135^\circ$ and $170^\circ$. 
Additionally, the four asteroids (25) Phocaea, (1591) Baize, (4826) Wilhelms, and (24827) Maryphil form a 'center group' around $90^\circ$ in obliquity. 

Besides the 41 S-complex asteroids in our analysis, we also searched members of unique pole solutions from previous studies, and find other 131 Phocaea family members \citep{2013AA...559A.134H,2013AA...551A..67H,Hanus2016,2018Icar..309..297H,Hanus2021AA...654A..48H,Durech2016AA...587A..48D,Durech2018AA...617A..57D,2019AA...631A...2D,Durech2020AA...643A..59D,Durech2023,Cellino2024,2022PSJ.....3...56H}, those data are shown as gray dots in the left panel of Figure~\ref{Dist-pole}. Incorporating these additional data, By comparison, pole longitudes from literature show more equilibrated, while the peak of retrograde is more significant than our samples. The ratios of prograde and retrograde rotators for whole data set and our samples are 60:112 and 15:26 respectively.

The bimodal distribution of the obliquity with two peaks around $32^{\circ}$ and $155^{\circ}$ reflects the evidence of the long-term YORP effect on the pole orientation of the asteroids. This is consistent with the theoretical predictions \citep{Bottke2006,PAOLICCHI2016314,Vokrouhlicky2015} that obliquities of family members are changed toward extreme values $0^\circ$ and $180^\circ$ driven by the YORP effect. We conclude that the deviation of the two peaks from the theoretical extrema of $0^\circ$ and $180^\circ$ reflects the fact  that pole re-orientation is efficient near the theoretical extrema.  
In the analysis of the YORP effect for the Koronis members, \cite{Slivan2002} first suggested the existence of the so-called Slivan states, the alignment of spin vectors and correlation of spin rates. \cite{Kryszczynska2013} also claimed that several members in the Flora family stayed in the Slivan state. As a comparison, we display the obliquities vs spin rates of 41 Phocaea family members with red symbols, Koronis members with black symbols, and Flora members with green symbols in the right panel of Figure \ref{Dist-pole}. For the Koronis and Flora families, spin rates for the prograde rotators show a more significant cluster feature than that for the retrograde rotators. For the case of the Phocaea members, neither the prograde nor the retrograde rotators show a significant cluster trend in the spin rate distribution, whereas, the 'center group' shows a significant cluster with spin rates ranging from $1.6 \sim 3.1$ cycles/day. 
\begin{figure}[ht]
    \centering
    \includegraphics[width=0.47\textwidth]{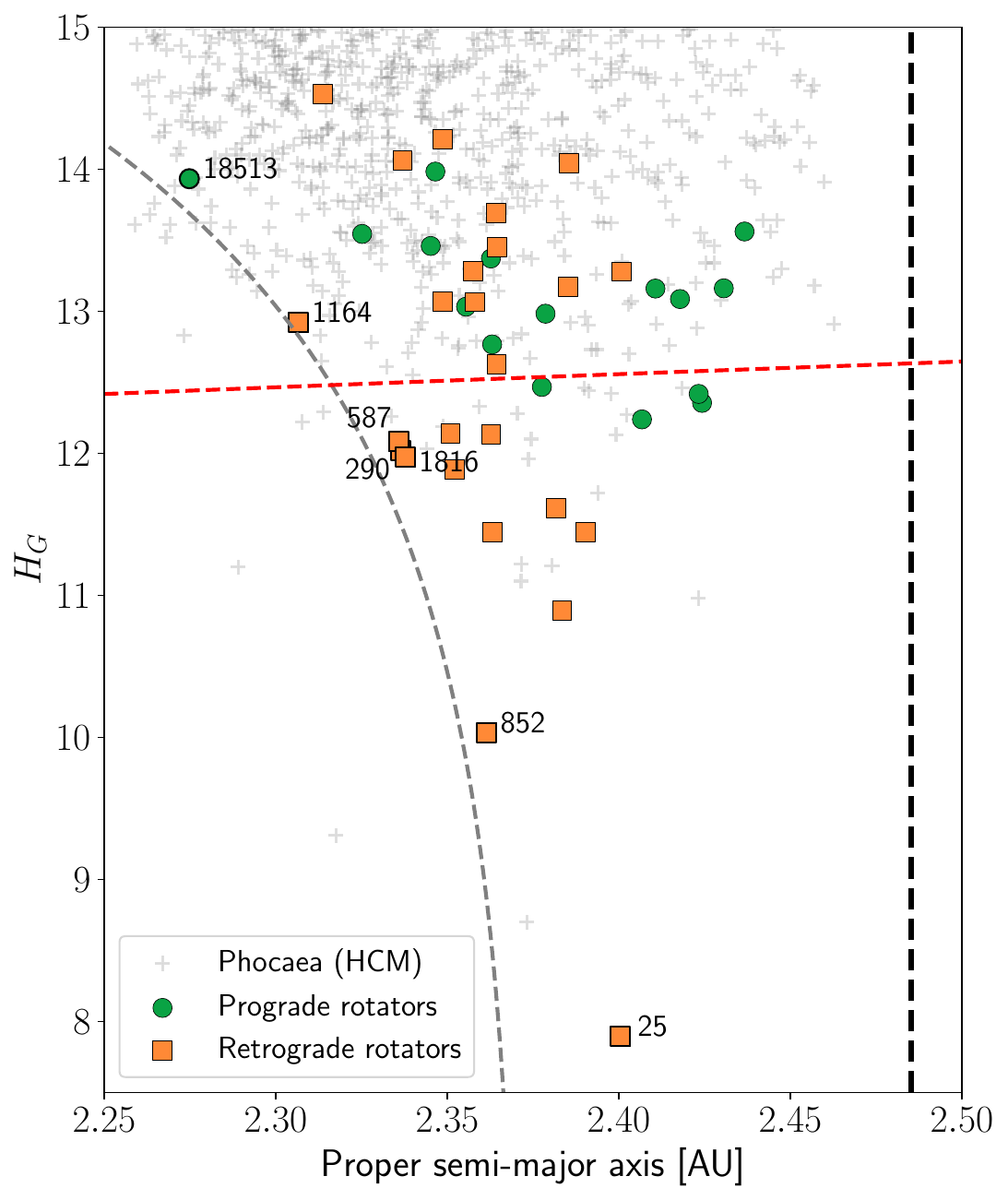}
    \caption{The distribution of the Phocaea family members in the ($H$,$a$) plane, our targets being marked with squares and circles for retrograde and prograde rotators, respectively. The gray line indicates the inner boundary of the V-shape of the Phocaea family, while the black dashed line represents the inner boundary of the 3:1 resonance with Jupiter. 
    }
    \label{Dist-pole2}
\end{figure}

\begin{deluxetable*}{rcrccrr}[ht]
    \setlength{\tabcolsep}{8pt}
    \tablecaption{Information on the rates of orbital drift ($(da/dt)_{\textit{orb}}$), asteroid diameters ($D$), Bond albedo ($A$), obliquity ($\gamma$), bulk density ($\rho$) and macroporosity ($p$) for five asteroids.}
    \label{density}
    \tablehead{
    \colhead{Asteroid} & \colhead{$(da/dt)_{\textit{orb}}$} & \colhead{$D$} & \colhead{$A$} & \colhead{$\gamma$} & \colhead{$\rho$} & \colhead{$p$} \\
    \colhead{} & \colhead{($10^{-4}/Myr$)} & \colhead{(km)} & \colhead{} & \colhead{deg} & \colhead{($g/cm^{3} $)} &
    \colhead{(\%)}
    }
    \startdata
        290 & $-0.300 \pm 0.029$ & $9.82\pm0.13$ & $0.133\pm0.025$ &$138.5\pm1.5$& $2.19 \pm 0.23$ & 41.2 \\ 
        587 & $-0.304 \pm 0.030$ & $11.02\pm0.086$ & $0.095\pm0.022$ &$120.2\pm1.7$& $1.29 \pm 0.15$ & 65.4 \\ 
        852 & $-0.088 \pm 0.008$ & $26.54\pm0.155$ & $0.103\pm0.006$ &$158.8\pm1.7$& $3.54 \pm 0.36$ &  4.8 \\ 
        1164 & $-0.551 \pm 0.054$ & $7.65\pm0.064$ & $0.091\pm0.011$ &$160.4\pm2.2$& $1.97 \pm 0.20$ &  47.1 \\ 
        1816 & $-0.288 \pm 0.028$ & $9.78\pm0.371$ & $0.126\pm0.008$ &$150.9\pm1.6$& $2.64 \pm 0.28$ & 29.2 \\
    \enddata
\end{deluxetable*}

The Figure \ref{Dist-pole2} shows a V-shape distribution of the $H_G$ magnitude vs semimajor axis ($a$) of the Phocaea family members (gray pluses for all members, circles and squares symbols for the 41 analyzed targets in this work). Theoretical studies \citep{VOKROUHLICKY2006,Vokrouhlicky2015,PAOLICCHI2016314} show that the formation of the V-shape is due to the combined influence of the Yarkovsky and YORP effects. Since the family was formed by a collisional event, the Yarkovsky effect drives the prograde and retrograde fragments outward and inward, respectively. The smaller members of the family tend to move farther away from the parent body, resulting in such a V-shape in the $(H_G,a)$ space over millions of years. Based on the V-shape of the Phocaea family, \cite{Milani2017} estimated its age of $1187\pm319$ Myr by using the inner boundary of the V-shape. The fuzzy outer boundary of the Phocaea family (see Figure \ref{Dist-pole2}) is due to the effect of the 3:1 resonance with Jupiter. 
In the V-shape plot, the prograde and retrograde rotators are distinguished by orange and green colors, respectively.
For our samples below the red line ($H_G = 12.4$~mag, $D = 8.8$~km assuming $p_{_G}$ = 0.25), retrograde rotators are strictly in the left part of the V-shape, and the prograde ones in the right part. For our samples above the red line, they mix to some extent. This mixture phenomenon could be the result of the YORP cycle. This refers to a timescale where the pole obliquity of an asteroid decreases or increases to $0^\circ$ or $180^\circ$, respectively, due to the YORP effect, whereafter the pole of the asteroid tends to be re-orientated randomly. Roughly, the YORP cycle is proportional to the sizes and heliocentric distances of the asteroids. Here, according to the method by \cite{PAOLICCHI2016314} and considering the Phocaea family age to be equal to the YORP cycle, we computed the H-threshold of the YORP cycle (drawn as the red dashed line in the V-shape plot). Consequently, those samples located above the red line should have undergone at least one YORP cycle, and their poles may have been re-orientated randomly. By the YORP cycle, we can comprehend why the prograde asteroid (18513) 1996 TS5 occurs near the inner boundary of the V-shape.

\subsection{Estimation of asteroid densities}

The age of an asteroid family is a key to understanding its origin and evolution. Considering certain evolutionary mechanisms, several different methods have been developed to determine the age of asteroid families. For  collisional families, like the Phocaea family, the methods based on the Yarkovsky and YORP chronology (i.e., \cite{Nesvorny2003, VOKROUHLICKY2006,SPOTO2015,Bottke2015b}) are considered to be the most precise ones at present. 

\cite{SPOTO2015} determine the age of asteroid families based on the values of the slope of the V-shape derived by iterative least squares fitting with interloper removal. The inverse slope is $\Delta a$, the accumulated change in orbital semimajor $a$ over the family age with unit $1/D$. Accordingly, the Yarkovsky drift rate $da/dt$ for family members near the inner and outer boundaries of the V-shape may be estimated by assuming that they keep moving inward or outward since the formation of the family.

Among the 41 Phocaea family asteroids, we choose five target, the asteroids (290) Bruna, (587)  Hypsipyle, (852) Wladilena, (1164) Kobolda, and (1816) Liberia close to the inner boundary, for the calculation of their $\Delta a$. As for the asteroid (18513) 1996 TS5, We do not include the asteroid (18513) 1996 TS5 in the sample, because of the complication of the drift by the YORP cycle. In order to determine the accumulated semimajor axis change, we estimate the center of the V-shape of the Phocaea family ($a_c$) as 2.372~au by using the $dw$-method, a border technique for V-shape identification \citep{Delbo2017,Delbo2019} by fixing the slope value of the inner borders of the V-shape given by \cite{Milani2017}. With the estimated $a_c$ and family age ($\tau_{age}$), the approximation of the drift rates for the semimajor axis $(da/dt)_{\textit{orb}}$ for the five asteroids are calculated by $\Delta a/ \tau_{age}$. Considering that their migration in the orbital semimajor axis has risen from the Yarkovsky effect, we figure out their densities due to the successful determination of their physical parameters, especially the Bond albedo ($A$) and pole obliquity ($\gamma$). For the goal, we use the Eq.~(\ref{drift}) in \cite{SPOTO2015}:
\begin{equation}
    \begin{split}
    \frac{da}{dt} &= (\frac{da}{dt})_{Bennu}\frac{\sqrt{a_{Bennu}}}{\sqrt{a}}\frac{1-e_{Bennu}^2}{1-e^2}\frac{D_{Bennu}}{D} \\
    &\quad\times\frac{\rho_{Bennu}}{\rho} \frac{\cos(\gamma)}{\cos(\gamma_{Bennu})}\frac{1-A}{1-A_{Bennu}},
    \label{drift}
    \end{split}
\end{equation}
where $D$ is the diameter of the asteroid retrieved from the NEOWISE survey, $\gamma$ is the derived pole obliquity, $\rho$ is the bulk density, and $A$ is the Bond albedo computed from the geometric albedo $p_{_G}$ and the parameters $G_1,G_2$ of the $H,\!G_1,\!G_2$ photometric function. The subscript 'Bennu' means the corresponding parameters of asteroid (101955) Bennu. The measured Yarkovsky drift rate $({da}/{dt})_{Bennu}$ from the paper of \citet{2021Icar..36914594F} and the involved parameters of Bennu come from \cite{NOLAN2013629} and \cite{CHESLEY2014}. The estimated densities of five asteroids (see the 6th column of Table \ref{density}) range from $1.29\sim3.54$ $g/cm^{3}$. The uncertainty of $(da/dt)_{\textit{orb}}$ is estimated by a propagation of error considering the uncertainty of the family age. The uncertainty of bulk density was evaluated through 100,000 Monte Carlo simulations, taking into account the uncertainties in the $(da/dt)_{\textit{orb}}$, asteroid diameter, obliquity, and Bond albedo.

According to the spectroscopic data of the asteroids, the S-complex asteroids are linked to the ordinary chondrite meteorites  (OCs). Compared to densities of OCs, the density of asteroid (852) Wladilena is very close to that of the H chondrites, which is consistent with the result of \cite{Noonan2019}, that is, the core member (25) Phocaea is associated to H chondrites.

Using the average grain density of H chondrites $\rho_m=3.72$ $g/cm^{3}$, the macroporosity of the five asteroids are calculated with the formula $P(\%)=100(1-\rho/\rho_m)$ \citep{CARRY201298}, as shown in Table \ref{density}. From the macroporosity values, the asteroids (290) Bruna, (587) Hypsipyle, and (1164) Kobolda have high macroporosities, indicating rubble pile structures. 
Considering the shapes of (290) Bruna, (587) Hypsipyle, and (1164) Kobolda, we find them to resemble 'top' shape (see Figure \ref{shape}).

\section{Summary}
(1) Combining dense and sparse photometric data obtained by ground-based and space-based telescopes, the physical parameters (including shape, spin parameters, and phase function parameters) of 44 asteroids in the Phocaea region are determined with the Bayesian lightcurve inversion method. 

(2) We derived the absolute Gaia magnitude $H_G$, the photometric slope $\beta_{\textit{ref}}$, and the geometric albedo in the G band $p_{_G}$ for the 44 asteroids. The estimated slopes and geometric albedos of 41 asteroids reflect the S-complex composition and that of 3 asteroids (105) Artemis, (1942) Jablunka, and (1963) Bezovec are C-complex. The three asteroids are probable interlopers for the Phocaea family, while (105) Artemis and (1963) Bezovec do not belong to members of the Tamara family, because of their significantly larger sizes.

(3) From the distribution of pole longitude $\lambda_p$ vs pole obliquity $\gamma$ of the 41 S-complex members, a bimodal distribution in $\gamma$ is clear. The two existing clusters around the peak values of $32^\circ$  and $155^\circ$ are footprints of the long-term influence of the YORP effect on the Phocaea family.  As for the spin rate of 41 asteroids, no significant Slivan state can be found among the prograde and retrograde clusters. Instead, four asteroids with obliquity around $90^\circ$ have spin rates between $1.6 \sim 3.1$ cycles/day.  

(4) The distribution of the 41 asteroids in the V-shaped absolute magnitude vs. semimajor axis ($H_G$,$a$) plot of the Phocaea family also show a significant footprint of the Yarkovsky effect. In agreement with the theoretical predictions, retrograde rotators tend to be left in the V-shape and prograde rotators right except for a few samples. By using a threshold magnitude of $H_G = 12.4$~mag (equivalent to $D = 8.8$~km, assuming $p_{_G} = 0.25$) and the YORP cycle, the mixture of the prograde and retrograde rotators in the V-shape plot can be explained. The ratio of prograde to retrograde rotators is 0.57, which can be explained by the influence of the 3:1 mean motion resonance with Jupiter, due to which prograde asteroids tend to escape from the Phocaea region and may become sources of prograde rotating near-Earth asteroids. 

(5) By fitting the center position of the V-shape and family age, we estimate the drift rates of the semimajor axis $(da/dt)_{\textit{orb}}$ for five asteroids with retrograde rotation very close to the inner boundary of the V-shape. Assuming that the accumulated $\Delta a$ derives from the Yarkovsky effect in the time of the family age, we estimate the densities of the five asteroids. Among the asteroids, (852) Wladilena has the highest density of 3.54 $g/cm^{3}$ which is very close to the bulk density of H chondrite meteorites. Besides asteroid (25) Phocaea, this is an other member of H Chondrite material in Phocaea family. We think the parent body of the Phocaea family has been composed of H chondrite like material and the Phocaea family may be one of the sources of H chondrite meteorites.  (587) Hypsipyle has the lowest bulk density of 1.29 $g/cm^{3}$ which is similar to that of asteroid Bennu (1.26 $g/cm^{3}$, estimated by \cite{CHESLEY2014}). With the help of the grain density of H chondrites, we calculate the macroporosity of the five asteroids. The resulting macroporosity values of 41\%, 65\% , and 47\% for (290) Bruna, (587) Hypsipyle, and (1164) Kobolda imply a rubble-pile structure. Their top-like shapes support the conclusion.  

(6) Presently, a small number of 41 members of the Phocaea family are incorporated in the physical study of the family. Nevertheless, the footprints of the Yarkovsky and YORP effects in the Phocaea family can be clearly seen with the help of the members' physical parameters, that is, their shapes, spin parameters, and phase function parameters. The future ground-based and space-based surveys will dramatically improve the knowledge of asteroid families. For example, the ten-year survey of the Chinese Space Station Telescope (CSST) will provide 7-band photometric data (apparent limiting magnitude of 26 mag in the g-band) and slitless spectra (wavelength range $225 \sim 1000$ nm, apparent limiting magnitude of 23 mag) for faint members of high-inclination asteroid families, which can be used to determine the physical parameters of small family members. The present  work marks the beginning of the physical studies of asteroid families by using incoming data gathered by from both space-based and ground-based surveys.

\section*{Acknowledgments}
We are grateful to the referee for the useful suggestions and comments to the manuscript, which help to improve its contents. We would like to thank the financial support from the National Natural Science Foundation of China under grant No.12288102 and the Yunnan Fundamental Research Project (grant No. 202305AS350009). This work is also supported by the National Natural Science Foundation of China (grants No. 12373069 ) and the Research Council of Finland (grants No. 359893 and 336546). We acknowledge the science research grants from the China Manned Space Project with No. CMS-CSST-2025-A16 ,CMS-CSST-2025-A18 and CMS-CSST-2025-A20, the Foreign Experts Project (FEP) State Administration of Foreign Experts Affairs of China (SAFEA) with No.~G2021039001L and H20240864, and the Chinese Academy of Sciences President’s International Fellowship Initiative (PIFI) Grant No.~2021VMA0017.

This work uses data downloaded  from the Asteroid Lightcurve Data Exchange Format (ALCDEF) database, which is supported by funding from the NASA grant 80NSSC18K0851. We also used data collected by the TESS mission funded by the NASA’s Science Mission Directorate. This work has made use of data from the European Space Agency mission {\it Gaia} (\url{https://www.cosmos.esa.int/gaia}), processed by the {\it Gaia} Data Processing and Analysis Consortium (DPAC, \url{https://www.cosmos.esa.int/web/gaia/dpac/consortium}). 
Funding for the DPAC has been provided by national institutions, in particular the institutions participating in the {\it Gaia} Multilateral Agreement.

\clearpage
\appendix
\section{Appendix A: Physical parameters of 44 asteroids in the Phocaea region}
Table \ref{Information on spin parameters} summarizes the spin parameters of 44 asteroids determined in this work and the previous work.
\startlongtable
\begin{deluxetable}{lllllrrrrrrrrr}
    \tabletypesize{\scriptsize}
    \renewcommand{\arraystretch}{0.85}
    \setlength{\tabcolsep}{1.4pt}
    \tablecaption{Spin parameters of 44 asteroids in the Phocaea region.}\label{Information on spin parameters}
    \tablehead{
    \colhead{Asteroid} & 
    \multicolumn{4}{c}{Rotation period} & 
    \multicolumn{4}{c}{Pole orientation (this work)} & 
    \multicolumn{4}{c}{Reference pole} & 
    \colhead{Ref.} \\
    &
    \colhead{$P$} & 
    \colhead{$\sigma_{_\textit{p}}$} &
    \colhead{$P_{\textit{ref}}$} & 
    \colhead{$\sigma_{_{\textit{p,ref}}}$} &
    \colhead{$\lambda_1$} & 
    \colhead{$\beta_1$} & 
    \colhead{$\lambda_2$} & 
    \colhead{$\beta_2$} & 
    \colhead{$\lambda_1$} & 
    \colhead{$\beta_1$} & 
    \colhead{$\lambda_2$} & 
    \colhead{$\beta_2$} & 
    \colhead{}\\
    &
    \colhead{(hr)} & 
    \colhead{($10^{-5}$ hr)} & 
    \colhead{(hr)} & 
    \colhead{($10^{-5}$ hr)} & 
    \colhead{$(^\circ)$} &
    \colhead{$(^\circ)$} &
    \colhead{$(^\circ)$} &
    \colhead{$(^\circ)$} &
    \colhead{$(^\circ)$} &
    \colhead{$(^\circ)$} &
    \colhead{$(^\circ)$} &
    \colhead{$(^\circ)$} &
    }
    \startdata\\
        25 & $9.9354075$ & $1.65$ & 9.9354 & 1.0 & $344.2\pm0.6$ & $9.6\pm0.9$ & ~ & ~ & $347\pm10$ & $10\pm10$ & ~ & ~ & \tablenotemark{\scriptsize{a}} \\ \\
        105 & $37.1197492$ & $3.86$ & 37.118 & ~ & $44.7\pm1.4$ & $16.6\pm1.4$ & ~ & ~ & 47 & 24 & ~ & ~ & \tablenotemark{\scriptsize{g}} \\ 
        ~ & ~ & ~ & $37.16$ & 1000.0 & ~ & ~ & ~ & ~ & $233.5\pm5$ & $-42.5\pm5$ & ~ & ~ & \tablenotemark{\scriptsize{h}} \\ \\
        290 & $13.8055624$ & $0.89$ & 13.80554 & 1.0 & $259.1\pm1.5$ & $-70.1\pm1.4$ & ~ & ~ & $286\pm10$ & $-80\pm10$ & $37\pm10$ & $-74\pm10$ & \tablenotemark{\scriptsize{a}} \\ \\
        502 & $10.9266526$ & $0.56$ & 10.92666 & ~ & $174.5\pm0.7$ & $-41.3\pm0.6$ & ~ & ~ & $178\pm6$ & $-36\pm5$ & ~ & ~ & \tablenotemark{\scriptsize{b}} \\ 
        ~ & ~ & ~ & 10.92666 & ~ & ~ & ~ & ~ & ~ & $184$ & $-39$ & 285 & -45 & \tablenotemark{\scriptsize{e}} \\ \\
        587 & $2.8894654$ & $0.39$ & $2.889463$ & 0.7 & $173.7\pm1.8$ & $-45.5\pm4.8$ & $323.4\pm3.2$ & $-31.6\pm3.0$ & $167\pm4$ & $-60\pm4$ & ~ & ~ & \tablenotemark{\scriptsize{d}} \\ 
        ~ & ~ & ~ & $13.6816$ & 50.0 & ~ & ~ & ~ & ~ & $232\pm18$ & $36\pm15$ & $55\pm18$ & $32\pm15$ & \tablenotemark{\scriptsize{b}} \\ \\
        852 & $4.6132965$ & $0.08$ & 4.613301 & 1.0 & $160.6\pm2.2$ & $-58.2\pm0.7$ & ~ & ~ & $181\pm10$ & $-48\pm10$ & $46\pm10$ & $-53\pm10$ & \tablenotemark{\scriptsize{a}} \\ 
        ~ & ~ & ~ & 4.61334 & 1.0 & ~ & ~ & ~ & ~ & $170\pm20$ & $-57\pm20$ & ~ & ~ & \tablenotemark{\scriptsize{f}} \\ 
        ~ & ~ & ~ & 4.6133 & ~ & ~ & ~ & ~ & ~ & 45 & -53 & ~ & ~ & \tablenotemark{\scriptsize{e}} \\ \\
        1164 & $4.1417002$ & $0.06$ & $4.14168$ & 2.0 & $278.6\pm2.6$ & $-53.3\pm0.7$ & ~ & ~ & $271\pm9$ & $-60\pm3$ & $6\pm6$ & $-54\pm8$ & \tablenotemark{\scriptsize{d}} \\ \\
        1192 & $6.5583613$ & $0.02$ & 6.55836 & ~ & $160.5\pm0.9$ & $-79.1\pm0.3$ & ~ & ~ & $133\pm24$ & $-78\pm5$ & $268\pm16$ & $-73\pm5$ & \tablenotemark{\scriptsize{b}} \\ 
        ~ & ~ & ~ & $6.55836$ & 1.0 & ~ & ~ & ~ & ~ & $144\pm20$ & $-66\pm20$ & ~ & ~ & \tablenotemark{\scriptsize{f}} \\ \\
        1568 & $6.6759829$ & $0.29$ & 6.67597 & ~ & $121.4\pm1.8$ & $-67.5\pm0.7$ & ~ & ~ & 109 & -68 & ~ & ~ & \tablenotemark{\scriptsize{i}} \\ 
        ~ & ~ & ~ & 6.67601 & ~ & ~ & ~ & ~ & ~ & 142 & -75 & 279 & -106 & \tablenotemark{\scriptsize{e}} \\ \\
        1591 & $7.7944596$ & $0.69$ & $7.79347$ & 7.0 & $193.1\pm0.7$ & $23.7\pm1.1$ & ~ & ~ & $189\pm1$ & $45\pm6$ & $5\pm2$ & $-2\pm4$ & \tablenotemark{\scriptsize{d}} \\ \\
        1626 & $3.4201645$ & $0.05$ & $3.42015$ & 0.7 & $251.7\pm1.5$ & $-43.2\pm1.4$ & ~ & ~ & $250\pm3$ & $-37\pm4$ & ~ & ~ & \tablenotemark{\scriptsize{d}} \\ 
        ~ & ~ & ~ & $3.421367$ & 0.3 & ~ & ~ & ~ & ~ & $152\pm10$ & $-9\pm10$ & ~ & ~ & \tablenotemark{\scriptsize{j}} \\ \\
        1816 & $3.0861619$ & $0.10$ & 3.086156 & ~ & $87.5\pm1.7$ & $-86.8\pm0.8$ & ~ & ~ & $73\pm26$ & $-68\pm10$ & ~ & ~ & \tablenotemark{\scriptsize{b}} \\ 
        ~ & ~ & ~ & $3.086156$ & 0.5 & ~ & ~ & ~ & ~ & $218\pm20$ & $-83\pm20$ & ~ & ~ & \tablenotemark{\scriptsize{f}} \\ \\
        1884 & $2.8944232$ & $0.54$ & $2.89442$ & 2.0 & $39.8\pm1.0$ & $50.0\pm0.7$ & ~ & ~ & $42\pm3$ & $52\pm3$ & ~ & ~ & \tablenotemark{\scriptsize{d}} \\ \\
        1942 & $8.9116186$ & $1.44$ & $8.91158$ & 1.0 & $187.0\pm2.6$ & $-79.6\pm0.5$ & ~ & ~ & $156\pm15$ & $-73\pm15$ & ~ & ~ & \tablenotemark{\scriptsize{p}} \\ 
        ~ & ~ & ~ & $8.9115$ & ~ & ~ & ~ & ~ & ~ & $220\pm20$ & $-55\pm20$ & ~ & ~ & \tablenotemark{\scriptsize{f}} \\ \\
        1963 & $18.1654180$ & $0.71$ & 18.1655 & 1.0 & $221.5\pm0.6$ & $1.4\pm1.1$ & ~ & ~ & $219\pm10$ & $7\pm10$ & ~ & ~ & \tablenotemark{\scriptsize{a}} \\ 
        ~ & ~ & ~ & 18.1655 & 10.0 & ~ & ~ & ~ & ~ & $223\pm20$ & $-7\pm20$ & ~ & ~ & \tablenotemark{\scriptsize{f}} \\ 
        ~ & ~ & ~ & 18.16539 & ~ & ~ & ~ & ~ & ~ & 222 & 2 & 73 & -65 & \tablenotemark{\scriptsize{e}} \\ \\
        1987 & $9.4594916$ & $0.12$ & 9.4595 & 1.0 & $349.2\pm1.7$ & $-59.3\pm0.8$ & ~ & ~ & $352\pm10$ & $-58\pm10$ & ~ & ~ & \tablenotemark{\scriptsize{a}} \\ 
        ~ & ~ & ~ & $9.4595$ & 2.0 & ~ & ~ & ~ & ~ & $352\pm10$ & $-52\pm10$ & ~ & ~ & \tablenotemark{\scriptsize{k}} \\ \\
        2430 & $129.7460092$ & $51.91$ & 129.75 & 1000.0 & $176.7\pm1.7$ & $-68.2\pm0.8$ & ~ & ~ & $177\pm5$ & $-68\pm5$ & ~ & ~ & \tablenotemark{\scriptsize{l}} \\ \\
        2830 & $80.5771088$ & $45.18$ & $80.573$ & 200.0 & $321.2\pm1.3$ & $44.0\pm0.7$ & ~ & ~ & $315\pm5$ & $34\pm5$ & ~ & ~ & \tablenotemark{\scriptsize{m}} \\ \\
        3388 & $3.2575139$ & $0.14$ & $3.25748$ & 1.0 & $97.2\pm2.9$ & $82.4\pm0.8$ & ~ & ~ & $91\pm20$ & $82\pm20$ & ~ & ~ & \tablenotemark{\scriptsize{f}} \\ \\
        3895 & $3.5646214$ & $0.04$ & 3.56468 & 1.0 & $34.1\pm3.9$ & $-84.0\pm1.0$ & ~ & ~ & $33\pm20$ & $-74\pm20$ & $233\pm20$ & $-22\pm20$ & \tablenotemark{\scriptsize{f}} \\ 
        ~ & ~ & ~ & 3.56462 & ~ & ~ & ~ & ~ & ~ & 49 & -84 & 236 & -15 & \tablenotemark{\scriptsize{e}} \\ 
        ~ & ~ & ~ & 3.56461 & ~ & ~ & ~ & ~ & ~ & $349\pm4$ & $-86\pm1$ & ~ & ~ & \tablenotemark{\scriptsize{d}} \\ \\
        4132 & $3.2963327$ & $0.08$ & $3.29633$ & 1.0 & $279.4\pm1.9$ & $71.2\pm1.7$ & ~ & ~ & 281 & 72 & ~ & ~ & \tablenotemark{\scriptsize{a}} \\ 
        ~ & ~ & ~ & 3.296333 & ~ & ~ & ~ & ~ & ~ & 186 & 87 & ~ & ~ & \tablenotemark{\scriptsize{c}} \\ \\
        4742 & $5.4810471$ & $0.55$ & $5.48105$ & 1.0 & $173.3\pm1.1$ & $45.3\pm1.0$ & ~ & ~ & $175\pm20$ & $37\pm20$ & ~ & ~ & \tablenotemark{\scriptsize{f}} \\ 
        4826 & $15.0043414$ & $0.82$ & 15.0043 & ~ & $299.4\pm0.3$ & $1.7\pm1.0$ & ~ & ~ & $300\pm3$ & $3\pm5$ & ~ & ~ & \tablenotemark{\scriptsize{d}} \\ 
        ~ & ~ & ~ & 15.00444 & ~ & ~ & ~ & ~ & ~ & 298 & 3 & ~ & ~ & \tablenotemark{\scriptsize{e}} \\ \\
        5040 & $4.6901545$ & $0.11$ & $4.690155$ & 0.2 & $94.6\pm1.6$ & $42.7\pm0.9$ & ~ & ~ & $101\pm10$ & $48\pm10$ & ~ & ~ & \tablenotemark{\scriptsize{n}} \\ \\
        5325 & $4.0235788$ & $0.12$ & $4.02356$ & 1.0 & $267.5\pm3.1$ & $65.7\pm1.6$ & ~ & ~ & $286\pm15$ & $73\pm15$ & ~ & ~ & \tablenotemark{\scriptsize{o}} \\ \\
        5647 & $6.1387299$ & $0.87$ & 6.1386 & ~ & $258.3\pm1.9$ & $64.0\pm0.8$ & ~ & ~ & $263\pm20$ & $51\pm20$ & ~ & ~ & \tablenotemark{\scriptsize{f}} \\ 
        ~ & ~ & ~ & 6.13867 & 1.0 & ~ & ~ & ~ & ~ & $266$ & $69$ & ~ & ~ & \tablenotemark{\scriptsize{a}} \\ 
        ~ & ~ & ~ & 6.13813 & ~ & ~ & ~ & ~ & ~ & 115 & 13 & ~ & ~ & \tablenotemark{\scriptsize{e}} \\ \\
        6510 & $6.3649158$ & $0.36$ & $6.3649$ & 1.0 & $91.5\pm2.2$ & $-84.3\pm2.6$ & $273.2\pm1.1$ & $-79.3\pm1.8$ & $84\pm15$ & $-72\pm15$ & $249\pm15$ & $-37\pm15$ & \tablenotemark{\scriptsize{p}} \\ \\
        6522 & $7.6942197$ & $0.84$ & 7.69424 & ~ & $353.4\pm2.3$ & $-67.2\pm0.4$ & ~ & ~ & 353 & -66 & ~ & ~ & \tablenotemark{\scriptsize{c}} \\
        ~ & ~ & ~ & $7.6943$ & 10.0 & ~ & ~ & ~ & ~ & $331\pm20$ & $-61\pm20$ & ~ & ~ & \tablenotemark{\scriptsize{f}} \\ \\
        6560 & $19.1980260$ & $2.01$ & $19.1981$ & 10.0 & $174.2\pm1.0$ & $51.4\pm2.3$ & ~ & ~ & $174\pm20$ & $48\pm20$ & ~ & ~ & \tablenotemark{\scriptsize{f}} \\ \\
        7055 & $4.1687842$ & $0.04$ & $4.16879$ & 1.0 & $148.1\pm2.5$ & $-62.4\pm0.6$ & ~ & ~ & $112\pm20$ & $-88\pm20$ & ~ & ~ & \tablenotemark{\scriptsize{f}} \\ \\
        8356 & $3.0432333$ & $0.05$ & ~ & ~ & $63.8\pm1.7$ & $74.4\pm1.3$ & ~ & ~ & ~ & ~ & ~ & ~ & ~ \\ \\
        8893 & $22.2785880$ & $6.46$ & $22.278$ & 100.0 & $237.5\pm1.7$ & $-76.2\pm1.8$ & ~ & ~ & $225\pm20$ & $-78\pm20$ & ~ & ~ & \tablenotemark{\scriptsize{f}} \\ \\
        11271 & $6.3616731$ & $1.19$ & $6.36167$ & 1.0 & $122.8\pm1.7$ & $-42.6\pm1.8$ & ~ & ~ & 104 & -29 & 260 & -29 & \tablenotemark{\scriptsize{c}} \\ \\
        15779 & $19.5284442$ & $6.60$ & $19.5268$ & 10.0 & $144.4\pm1.4$ & $38.1\pm1.6$ & ~ & ~ & $157\pm20$ & $51\pm20$ & ~ & ~ & \tablenotemark{\scriptsize{f}} \\ \\
        18513 & $9.7459095$ & $0.42$ & $9.7458$ & 10.0 & $188.3\pm1.1$ & $41.6\pm0.8$ & ~ & ~ & $194\pm20$ & $38\pm20$ & ~ & ~ & \tablenotemark{\scriptsize{f}} \\ \\
        22275 & $40.4764058$ & $6.54$ & $40.478$ & 100.0 & $151.8\pm2.3$ & $-81.1\pm0.9$ & ~ & ~ & $127\pm20$ & $-76\pm20$ & ~ & ~ & \tablenotemark{\scriptsize{f}} \\ \\
        23200 & $16.3061677$ & $0.85$ & $16.3061$ & 10.0 & $191.9\pm1.0$ & $-50.4\pm1.1$ & ~ & ~ & $199\pm20$ & $-56\pm20$ & ~ & ~ & \tablenotemark{\scriptsize{f}} \\ \\
        23552 & $3.6286870$ & $0.09$ & ~ & ~ & $321.0\pm4.6$ & $-73.2\pm0.9$ & $164.0\pm2.2$ & $-83.1\pm0.7$ & ~ & ~ & ~ & ~ & ~ \\ \\
        24827 & $11.6481124$ & $0.80$ & $11.6484$ & 50.0 & $349.0\pm0.5$ & $24.6\pm1.1$ & ~ & ~ & $354\pm7$ & $12\pm16$ & ~ & ~ & \tablenotemark{\scriptsize{d}} \\ \\
        29032 & $5.2372338$ & $0.15$ & $5.23725$ & 1.0 & $224.3\pm0.5$ & $28.5\pm0.6$ & ~ & ~ & $228\pm20$ & $41\pm20$ & ~ & ~ & \tablenotemark{\scriptsize{f}} \\ \\
        29729 & $4.5279868$ & $0.11$ & $4.52796$ & 1.0 & $165.4\pm0.7$ & $-89.1\pm0.6$ & $343.5\pm0.4$ & $-85.6\pm0.9$ & $165\pm20$ & $-89\pm20$ & ~ & ~ & \tablenotemark{\scriptsize{f}} \\ \\
        32036 & $4.7746088$ & $0.37$ & $4.7746$ & 10.0 & $130.2\pm0.6$ & $-83.7\pm0.6$ & ~ & ~ & $111\pm20$ & $-84\pm20$ & ~ & ~ & \tablenotemark{\scriptsize{f}} \\ \\
        54443 & $2.7993770$ & $0.14$ & $2.79937$ & 1.0 & $227.3\pm0.9$ & $-66.3\pm0.2$ & ~ & ~ & $222\pm20$ & $-65\pm20$ & ~ & ~ & \tablenotemark{\scriptsize{f}} \\ \\
        56086 & $2.7701508$ & $0.17$ & $2.77018$ & 1.0 & $44.6\pm3.7$ & $71.5\pm2.0$ & ~ & ~ & $13\pm20$ & $84\pm20$ & ~ & ~ & \tablenotemark{\scriptsize{f}} \\ \\
    \enddata
    \tablecomments{
    References: 
    \tablenotemark{\scriptsize{a}}~\citet{2013AA...551A..67H}; 
    \tablenotemark{\scriptsize{b}}~\citet{Hanus2016}; 
    \tablenotemark{\scriptsize{c}}~\citet{2019AA...631A...2D}; 
    \tablenotemark{\scriptsize{d}}~\citet{Durech2020AA...643A..59D}; 
    \tablenotemark{\scriptsize{e}}~\citet{2022PSJ.....3...56H}; 
    \tablenotemark{\scriptsize{f}}~\citet{Durech2023}; 
    \tablenotemark{\scriptsize{g}}~\citet{Hanus2021AA...654A..48H}; 
    \tablenotemark{\scriptsize{h}}~\citet{2008MPBu...35...63H}; 
    \tablenotemark{\scriptsize{i}}~\citet{2011AA...530A.134H}; 
    \tablenotemark{\scriptsize{j}}~\citet{Stephens2021MPBu...48...56S}; 
    \tablenotemark{\scriptsize{k}}~\citet{2018Icar..309..297H}; 
    \tablenotemark{\scriptsize{l}}~\citet{2013AA...559A.134H}; 
    \tablenotemark{\scriptsize{m}}~\citet{Durech2018AA...617A..57D}; 
    \tablenotemark{\scriptsize{n}}~\citet{2021MPBu...48..246S}; 
    \tablenotemark{\scriptsize{o}}~\citet{2016EMP..119...35H};
    \tablenotemark{\scriptsize{p}}~\citet{Durech2016AA...587A..48D}.
    }

\end{deluxetable}
\clearpage
Table \ref{Information on physical parameters} summarizes the phase function and related parameters of 44 asteroids determined in this work.
\begin{deluxetable}{lllrllrl}[h]
    \centering
    \renewcommand{\arraystretch}{1}
    \tabletypesize{\footnotesize}
    \tablewidth{0pt}
    \tablecaption{Phase function parameters and related parameters of 44 asteroids in the Phocaea region.}
    \label{Information on physical parameters}
    \tablehead{
    \colhead{Asteroid} & \colhead{$H_G$} & \colhead{$G_1$} & \colhead{$G_2$} &
    \colhead{$q$} & \colhead{$\beta_{\textit{ref}}$}  & \colhead{$D$} & \colhead{$p_{_G}$}\\
    \colhead{} & \colhead{(mag)} &  &  &  & \colhead{(mag/rad)}  & \colhead{km}&}
    \startdata
        25 & $7.90\pm0.09$ & $0.27\pm0.09$ & $0.43\pm0.10$ & $0.47\pm0.04$ & $1.55\pm0.12$ & $61.05\pm2.46$ & $0.33\pm0.03$ \\ 
        105 & $8.86\pm0.04$ & $0.80\pm0.21$ & $0.06\pm0.14$ & $0.38\pm0.02$ & $2.04\pm0.21$ & $94.86\pm23.22$ & $0.06\pm0.00$ \\ 
        290 & $12.02\pm0.13$ & $0.37\pm0.14$ & $0.38\pm0.15$ & $0.47\pm0.07$ & $1.54\pm0.17$ & $9.82\pm0.13$ & $0.28\pm0.04$ \\ 
        502 & $10.89\pm0.07$ & $0.39\pm0.06$ & $0.34\pm0.07$ & $0.45\pm0.04$ & $1.66\pm0.09$ & $15.80\pm1.16$ & $0.31\pm0.02$ \\ 
        587 & $12.08\pm0.14$ & $0.39\pm0.21$ & $0.34\pm0.20$ & $0.45\pm0.08$ & $1.68\pm0.23$ & $11.02\pm0.09$ & $0.21\pm0.03$ \\ 
        852 & $10.03\pm0.04$ & $0.43\pm0.03$ & $0.30\pm0.04$ & $0.43\pm0.02$ & $1.70\pm0.05$ & $26.54\pm0.15$ & $0.24\pm0.01$ \\ 
        1164 & $12.92\pm0.08$ & $0.36\pm0.13$ & $0.35\pm0.11$ & $0.44\pm0.04$ & $1.60\pm0.14$ & $7.65\pm0.06$ & $0.20\pm0.02$ \\ 
        1192 & $12.63\pm0.04$ & $0.50\pm0.03$ & $0.26\pm0.04$ & $0.42\pm0.02$ & $1.77\pm0.06$ & $7.38\pm0.19$ & $0.29\pm0.01$ \\ 
        1568 & $11.88\pm0.02$ & $0.49\pm0.06$ & $0.24\pm0.05$ & $0.41\pm0.01$ & $1.70\pm0.06$ & $12.45\pm0.08$ & $0.20\pm0.00$ \\ 
        1591 & $11.45\pm0.06$ & $0.42\pm0.10$ & $0.31\pm0.08$ & $0.43\pm0.03$ & $1.68\pm0.11$ & $14.41\pm0.24$ & $0.22\pm0.01$ \\ 
        1626 & $11.44\pm0.25$ & $0.30\pm0.22$ & $0.46\pm0.28$ & $0.51\pm0.14$ & $1.58\pm0.25$ & $15.14\pm0.49$ & $0.20\pm0.05$ \\ 
        1816 & $11.97\pm0.04$ & $0.44\pm0.08$ & $0.29\pm0.07$ & $0.42\pm0.02$ & $1.69\pm0.09$ & $9.78\pm0.37$ & $0.30\pm0.01$ \\ 
        1884 & $12.35\pm0.03$ & $0.38\pm0.10$ & $0.32\pm0.07$ & $0.43\pm0.02$ & $1.69\pm0.11$ & $8.42\pm0.39$ & $0.29\pm0.01$ \\ 
        1942 & $13.04\pm0.01$ & $0.96\pm0.07$ & $-0.06\pm0.05$ & $0.35\pm0.02$ & $2.27\pm0.06$ & $16.77\pm0.09$ & $0.04\pm0.00$ \\ 
        1963 & $11.09\pm0.01$ & $0.88\pm0.07$ & $-0.01\pm0.04$ & $0.36\pm0.01$ & $2.19\pm0.07$ & $35.54\pm0.23$ & $0.05\pm0.00$ \\ 
        1987 & $11.61\pm0.07$ & $0.39\pm0.06$ & $0.33\pm0.07$ & $0.44\pm0.03$ & $1.69\pm0.09$ & $13.02\pm0.16$ & $0.24\pm0.01$ \\ 
        2430 & $12.14\pm0.11$ & $0.32\pm0.08$ & $0.46\pm0.12$ & $0.51\pm0.06$ & $1.55\pm0.11$ & $12.47\pm0.32$ & $0.16\pm0.02$ \\ 
        2830 & $12.47\pm0.04$ & $0.41\pm0.07$ & $0.30\pm0.06$ & $0.42\pm0.02$ & $1.71\pm0.08$ & $7.89\pm0.10$ & $0.29\pm0.01$ \\ 
        3388 & $13.37\pm0.13$ & $0.27\pm0.12$ & $0.45\pm0.14$ & $0.48\pm0.07$ & $1.52\pm0.15$ & $6.29\pm0.31$ & $0.20\pm0.02$ \\ 
        3895 & $12.14\pm0.07$ & $0.41\pm0.05$ & $0.34\pm0.07$ & $0.45\pm0.03$ & $1.66\pm0.08$ & $10.38\pm0.09$ & $0.23\pm0.02$ \\ 
        4132 & $12.24\pm0.10$ & $0.34\pm0.11$ & $0.38\pm0.11$ & $0.45\pm0.05$ & $1.62\pm0.14$ & $9.72\pm0.06$ & $0.24\pm0.02$ \\ 
        4742 & $13.16\pm0.04$ & $0.56\pm0.06$ & $0.20\pm0.05$ & $0.40\pm0.02$ & $1.79\pm0.06$ & $6.55\pm0.14$ & $0.22\pm0.01$ \\ 
        4826 & $12.98\pm0.10$ & $0.43\pm0.09$ & $0.31\pm0.11$ & $0.44\pm0.05$ & $1.64\pm0.12$ & $7.93\pm0.12$ & $0.18\pm0.02$ \\ 
        5040 & $13.09\pm0.12$ & $0.32\pm0.13$ & $0.42\pm0.14$ & $0.48\pm0.06$ & $1.56\pm0.14$ & $7.06\pm0.00$ & $0.21\pm0.02$ \\ 
        5325 & $12.77\pm0.17$ & $0.35\pm0.25$ & $0.40\pm0.25$ & $0.48\pm0.11$ & $1.62\pm0.27$ & $6.97\pm0.23$ & $0.28\pm0.05$ \\ 
        5647 & $12.42\pm0.06$ & $0.36\pm0.04$ & $0.34\pm0.04$ & $0.43\pm0.02$ & $1.64\pm0.06$ & $8.61\pm0.23$ & $0.26\pm0.01$ \\ 
        6510 & $13.07\pm0.16$ & $0.33\pm0.11$ & $0.43\pm0.16$ & $0.49\pm0.08$ & $1.57\pm0.15$ & $6.66\pm0.09$ & $0.24\pm0.04$ \\ 
        6522 & $13.17\pm0.08$ & $0.42\pm0.08$ & $0.33\pm0.09$ & $0.45\pm0.04$ & $1.66\pm0.12$ & $6.12\pm0.12$ & $0.25\pm0.02$ \\ 
        6560 & $13.03\pm0.06$ & $0.37\pm0.05$ & $0.39\pm0.06$ & $0.47\pm0.03$ & $1.61\pm0.07$ & $7.60\pm0.00$ & $0.19\pm0.01$ \\ 
        7055 & $13.07\pm0.03$ & $0.41\pm0.02$ & $0.33\pm0.03$ & $0.44\pm0.02$ & $1.67\pm0.04$ & $6.69\pm0.65$ & $0.23\pm0.01$ \\ 
        8356 & $13.16\pm0.02$ & $0.56\pm0.11$ & $0.20\pm0.07$ & $0.40\pm0.02$ & $1.84\pm0.12$ & $6.75\pm0.31$ & $0.21\pm0.00$ \\ 
        8893 & $14.21\pm0.06$ & $0.52\pm0.12$ & $0.24\pm0.10$ & $0.41\pm0.03$ & $1.79\pm0.14$ & $4.12\pm0.08$ & $0.21\pm0.01$ \\ 
        11271 & $13.28\pm0.12$ & $0.43\pm0.11$ & $0.34\pm0.12$ & $0.45\pm0.06$ & $1.66\pm0.14$ & $6.82\pm0.07$ & $0.19\pm0.02$ \\ 
        15779 & $13.54\pm0.05$ & $0.38\pm0.06$ & $0.33\pm0.06$ & $0.43\pm0.03$ & $1.60\pm0.08$ & $5.70\pm0.10$ & $0.21\pm0.01$ \\ 
        18513 & $13.93\pm0.03$ & $0.42\pm0.05$ & $0.29\pm0.04$ & $0.42\pm0.02$ & $1.71\pm0.06$ & $4.11\pm0.69$ & $0.28\pm0.01$ \\ 
        22275 & $13.28\pm0.08$ & $0.37\pm0.10$ & $0.34\pm0.10$ & $0.44\pm0.04$ & $1.62\pm0.12$ & $6.05\pm0.01$ & $0.23\pm0.02$ \\ 
        23200 & $13.45\pm0.05$ & $0.32\pm0.05$ & $0.40\pm0.05$ & $0.46\pm0.03$ & $1.54\pm0.06$ & $5.29\pm1.17$ & $0.26\pm0.01$ \\ 
        23552 & $13.69\pm0.02$ & $0.46\pm0.06$ & $0.26\pm0.04$ & $0.40\pm0.01$ & $1.77\pm0.07$ & $5.46\pm0.00$ & $0.20\pm0.00$ \\ 
        24827 & $13.46\pm0.20$ & $0.42\pm0.21$ & $0.37\pm0.24$ & $0.48\pm0.11$ & $1.68\pm0.25$ & $5.95\pm0.35$ & $0.21\pm0.04$ \\ 
        29032 & $13.98\pm0.04$ & $0.50\pm0.05$ & $0.25\pm0.05$ & $0.41\pm0.02$ & $1.72\pm0.07$ & $4.16\pm0.05$ & $0.26\pm0.01$ \\ 
        29729 & $14.04\pm0.05$ & $0.40\pm0.06$ & $0.32\pm0.06$ & $0.43\pm0.03$ & $1.67\pm0.07$ & $4.31\pm0.03$ & $0.23\pm0.01$ \\ 
        32036 & $14.53\pm0.01$ & $0.44\pm0.04$ & $0.27\pm0.02$ & $0.41\pm0.01$ & $1.77\pm0.04$ & $3.56\pm0.53$ & $0.22\pm0.00$ \\ 
        54443 & $14.06\pm0.03$ & $0.31\pm0.03$ & $0.41\pm0.03$ & $0.47\pm0.01$ & $1.58\pm0.06$ & $3.73\pm0.19$ & $0.30\pm0.01$ \\ 
        56086 & $13.56\pm0.07$ & $0.29\pm0.06$ & $0.42\pm0.07$ & $0.47\pm0.03$ & $1.59\pm0.08$ & $5.24\pm0.09$ & $0.24\pm0.02$ \\
    \enddata
\end{deluxetable}
\clearpage

\section{Appendix B: Example asteroids for lightcurve inversion}
\begin{figure}[hb]
    \centering
    \begin{tabular}{c}
    \includegraphics[width=0.9\textwidth]{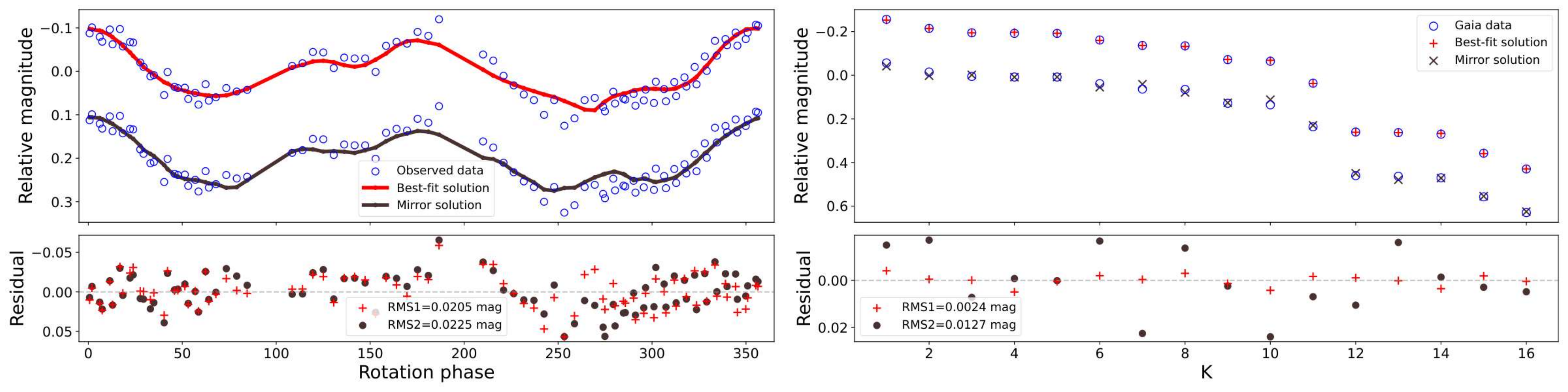}\\
    (a) Example ground-based photometric lightcurves \#3 and the Gaia lightcurves for asteroid (1626) Sadeya.\\
    \includegraphics[width=0.9\textwidth]{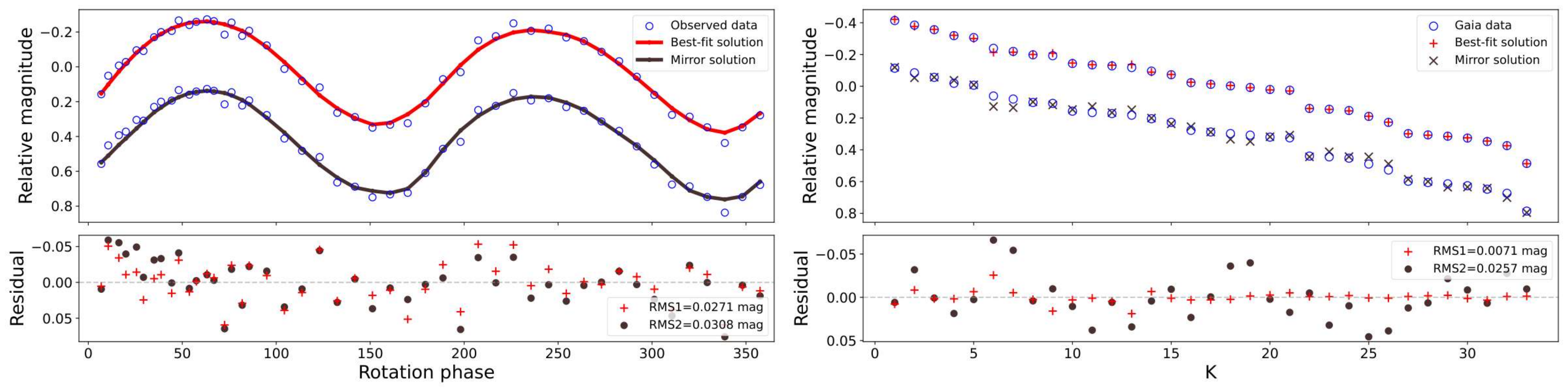}\\
    (b) Example TESS photometric lightcurves \#9 and the Gaia lightcurves for asteroid (6560) Pravdo.\\
    \includegraphics[width=0.9\textwidth]{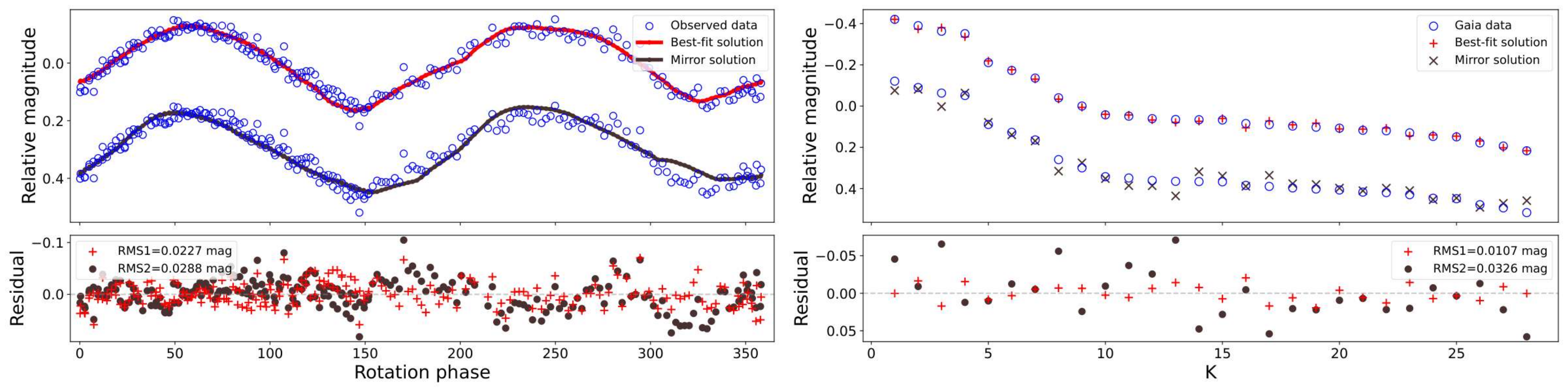}\\
    (c) Example ground-based photometric lightcurves \#10 and the Gaia lightcurves for asteroid (8356) Wadhwa.\\
    \includegraphics[width=0.9\textwidth]{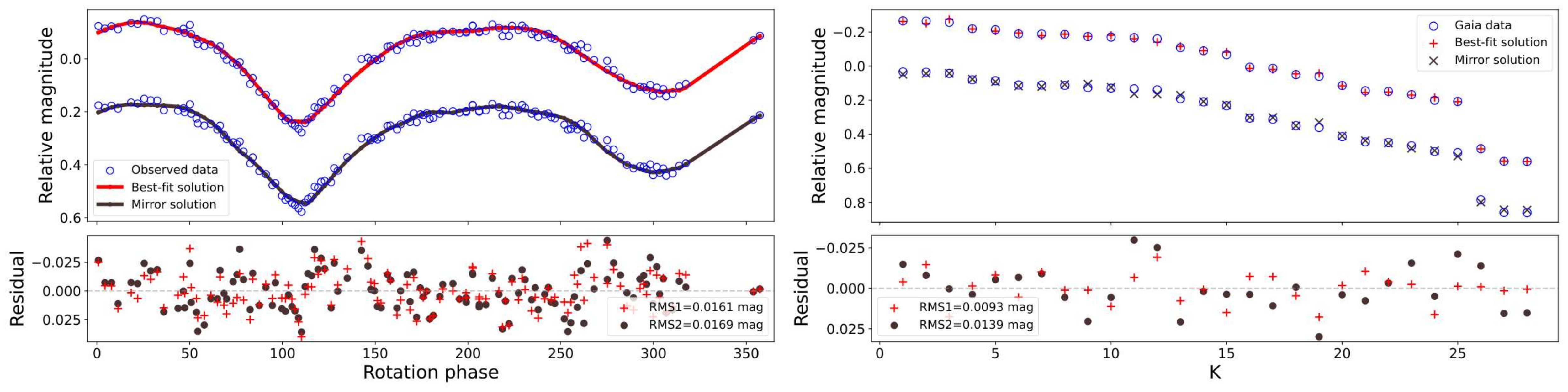}\\
    (d) Example ground-based photometric lightcurves \#5 and the Gaia lightcurves for asteroid (23552) 1994 NB.\\
    \end{tabular}
    \caption{
    Example lightcurves of four example asteroids, where K is the number of Gaia photometric data points.}  
    \label{lclc}
\end{figure}

Here, we present the dense lightcurves and Gaia photometric data of four example asteroids in Figure \ref{lclc}. The blue circles represent the observed data. The red and blue line represent the model magnitudes computed for the pairs of pole solutions. For clarity, lightcurves are offset for the two pole solutions. The bottom scatter plots describe 'observed minus computed' residuals for the observed and modeled photometric points, with red and black symbols representing the best-fit solution and its mirror solution, respectively.

\clearpage

\section{Appendix C: Convex shapes}
\begin{figure}[hb]
    \centering
    \includegraphics[width=0.9\textwidth]{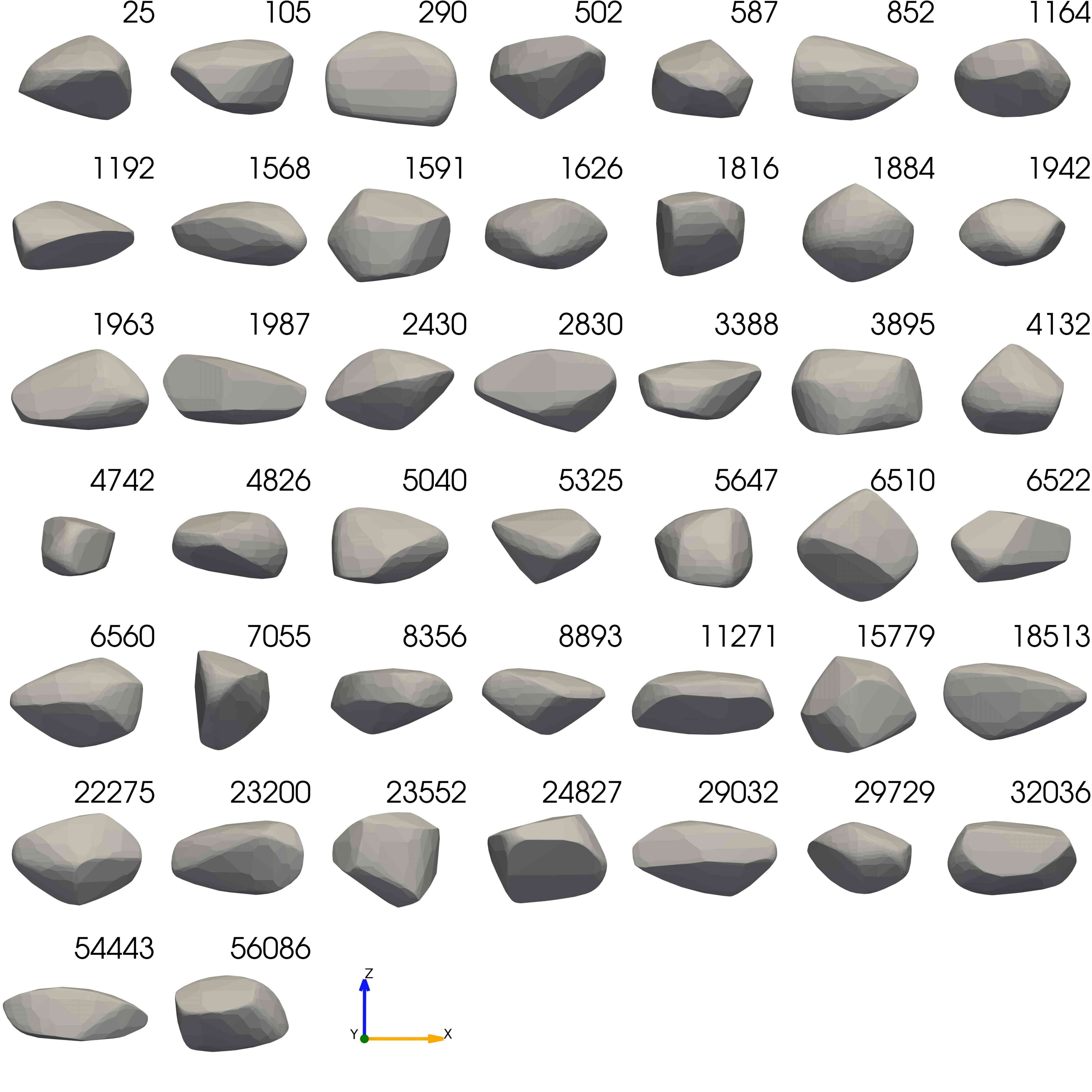}
    \caption{The convex shapes of 44 asteroids in the Phocaea region.}
    \label{shape}
\end{figure}

Figure \ref{shape} shows the convex shape of 44 asteroids corresponding to admissible pole solution, where the shapes of asteroids (587) Hypsipyle, (6510) Tarry, (23552) 1994 NB, and (29729) 1999 BY1 correspond to the pole solutions (173$^{\circ}$,-45$^{\circ}$), (91$^{\circ}$,84$^{\circ}$), (321$^{\circ}$,-73$^{\circ}$), and (343$^{\circ}$,-85$^{\circ}$), respectively.

\clearpage
\bibliography{sample631}{}

\begin{thebibliography}{}
\expandafter\ifx\csname natexlab\endcsname\relax\def\natexlab#1{#1}\fi
\providecommand{\url}[1]{\href{#1}{#1}}
\providecommand{\dodoi}[1]{doi:~\href{http://doi.org/#1}{\nolinkurl{#1}}}
\providecommand{\doeprint}[1]{\href{http://ascl.net/#1}{\nolinkurl{http://ascl.net/#1}}}
\providecommand{\doarXiv}[1]{\href{https://arxiv.org/abs/#1}{\nolinkurl{https://arxiv.org/abs/#1}}}

\bibitem[{{Bottke} {et~al.}(2015){Bottke}, {Bro{\v{z}}}, {O'Brien}, {Campo Bagatin}, {Morbidelli}, \& {Marchi}}]{Bottke2015b}
{Bottke}, W.~F., {Bro{\v{z}}}, M., {O'Brien}, D.~P., {et~al.} 2015, in Asteroids IV, ed. P.~Michel, F.~DeMeo, \& W.~Bottke (Tucson: University of Arizona Press), 701--724, \dodoi{10.2458/azu_uapress_9780816532131-ch036}

\bibitem[{Bottke {et~al.}(2000)Bottke, Jedicke, Morbidelli, Petit, \& Gladman}]{Bottke2000}
Bottke, W.~F., Jedicke, R., Morbidelli, A., Petit, J.-M., \& Gladman, B. 2000, Science, 288, 2190, \dodoi{10.1126/science.288.5474.2190}

\bibitem[{Bottke {et~al.}(2006)Bottke, Vokrouhlický, Rubincam, \& Nesvorný}]{Bottke2006}
Bottke, W.~F., Vokrouhlický, D., Rubincam, D.~P., \& Nesvorný, D. 2006, Annual Review of Earth and Planetary Sciences, 34, 157, \dodoi{https://doi.org/10.1146/annurev.earth.34.031405.125154}

\bibitem[{{Buchheim}(2009)}]{Buchheim2009MPBu...36...84B}
{Buchheim}, R.~K. 2009, Minor Planet Bulletin, 36, 84

\bibitem[{Bus \& Binzel(2002)}]{BUS2002146}
Bus, S.~J., \& Binzel, R.~P. 2002, Icarus, 158, 146, \dodoi{https://doi.org/10.1006/icar.2002.6856}

\bibitem[{Carruba(2009)}]{Carruba2009}
Carruba, V. 2009, Monthly Notices of the Royal Astronomical Society, 398, 1512, \dodoi{10.1111/j.1365-2966.2009.15220.x}

\bibitem[{Carry(2012)}]{CARRY201298}
Carry, B. 2012, Planetary and Space Science, 73, 98, \dodoi{https://doi.org/10.1016/j.pss.2012.03.009}

\bibitem[{Carvano {et~al.}(2001)Carvano, Lazzaro, Mothé-Diniz, Angeli, \& Florczak}]{Carvano2001}
Carvano, J.~M., Lazzaro, D., Mothé-Diniz, T., Angeli, C.~A., \& Florczak, M. 2001, Icarus, 149, 173, \dodoi{https://doi.org/10.1006/icar.2000.6512}

\bibitem[{Cellino {et~al.}(2024)Cellino, {Tanga, P.}, {Muinonen, K.}, \& {Mignard, F.}}]{Cellino2024}
Cellino, A., {Tanga, P.}, {Muinonen, K.}, \& {Mignard, F.} 2024, A\&A, 687, A277, \dodoi{10.1051/0004-6361/202449297}

\bibitem[{Chesley {et~al.}(2014)Chesley, Farnocchia, Nolan, Vokrouhlický, Chodas, Milani, Spoto, Rozitis, Benner, Bottke, Busch, Emery, Howell, Lauretta, Margot, \& Taylor}]{CHESLEY2014}
Chesley, S.~R., Farnocchia, D., Nolan, M.~C., {et~al.} 2014, Icarus, 235, 5, \dodoi{https://doi.org/10.1016/j.icarus.2014.02.020}

\bibitem[{Delbo {et~al.}(2019)Delbo, Avdellidou, \& Morbidelli}]{Delbo2019}
Delbo, M., Avdellidou, C., \& Morbidelli, A. 2019, A\&A, 624, A69, \dodoi{10.1051/0004-6361/201834745}

\bibitem[{Delbo {et~al.}(2017)Delbo, Walsh, Bolin, Avdellidou, \& Morbidelli}]{Delbo2017}
Delbo, M., Walsh, K., Bolin, B., Avdellidou, C., \& Morbidelli, A. 2017, Science, 357, 1026, \dodoi{10.1126/science.aam6036}

\bibitem[{DeMeo \& Carry(2013)}]{DEMEO2013}
DeMeo, F., \& Carry, B. 2013, Icarus, 226, 723, \dodoi{https://doi.org/10.1016/j.icarus.2013.06.027}

\bibitem[{{Durech} {et~al.}(2010){Durech}, {Sidorin}, \& {Kaasalainen}}]{Durech2010}
{Durech}, J., {Sidorin}, V., \& {Kaasalainen}, M. 2010, \aap, 513, A46, \dodoi{10.1051/0004-6361/200912693}

\bibitem[{{Farnocchia} {et~al.}(2021){Farnocchia}, {Chesley}, {Takahashi}, {Rozitis}, {Vokrouhlick{\'y}}, {Rush}, {Mastrodemos}, {Kennedy}, {Park}, {Bellerose}, {Lubey}, {Velez}, {Davis}, {Emery}, {Leonard}, {Geeraert}, {Antreasian}, \& {Lauretta}}]{2021Icar..36914594F}
{Farnocchia}, D., {Chesley}, S.~R., {Takahashi}, Y., {et~al.} 2021, \icarus, 369, 114594, \dodoi{10.1016/j.icarus.2021.114594}

\bibitem[{Granvik {et~al.}(2018)Granvik, Morbidelli, Jedicke, Bolin, Bottke, Beshore, Vokrouhlický, Nesvorný, \& Michel}]{GRANVIK2018}
Granvik, M., Morbidelli, A., Jedicke, R., {et~al.} 2018, Icarus, 312, 181, \dodoi{https://doi.org/10.1016/j.icarus.2018.04.018}

\bibitem[{Gu {et~al.}(2022)Gu, Wang, Yeung, Ng, Yu, Bai, Fan, Sun, Xiang, Cao, Lun, Xin, Wang, Xu, \& Liu}]{Gu2022}
Gu, S., Wang, X., Yeung, B., {et~al.} 2022, Astronomische Nachrichten, 343, e20224022, \dodoi{https://doi.org/10.1002/asna.20224022}

\bibitem[{{Hanu{\v{s}}} {et~al.}(2018){Hanu{\v{s}}}, {Delbo'}, {{\v{D}}urech}, \& {Al{\'\i}-Lagoa}}]{2018Icar..309..297H}
{Hanu{\v{s}}}, J., {Delbo'}, M., {{\v{D}}urech}, J., \& {Al{\'\i}-Lagoa}, V. 2018, Icarus, 309, 297, \dodoi{10.1016/j.icarus.2018.03.016}

\bibitem[{{Hanu{\v{s}}} {et~al.}(2021){Hanu{\v{s}}}, {Pejcha}, {Shappee}, {Kochanek}, {Stanek}, \& {Holoien}}]{Hanus2021AA...654A..48H}
{Hanu{\v{s}}}, J., {Pejcha}, O., {Shappee}, B.~J., {et~al.} 2021, Astronomy and Astrophysics, 654, A48, \dodoi{10.1051/0004-6361/202140759}

\bibitem[{{Hanu{\v{s}}} {et~al.}(2011){Hanu{\v{s}}}, {{\v{D}}urech}, {Bro{\v{z}}}, {Warner}, {Pilcher}, {Stephens}, {Oey}, {Bernasconi}, {Casulli}, {Behrend}, {Polishook}, {Henych}, {Lehk{\'y}}, {Yoshida}, \& {Ito}}]{2011AA...530A.134H}
{Hanu{\v{s}}}, J., {{\v{D}}urech}, J., {Bro{\v{z}}}, M., {et~al.} 2011, Astronomy and Astrophysics, 530, A134, \dodoi{10.1051/0004-6361/201116738}

\bibitem[{Hanu\v{s} {et~al.}(2013)Hanu\v{s}, {Bro\v{z}, M.}, {\v{D}urech, J.}, {Warner, B. D.}, {Brinsfield, J.}, {Durkee, R.}, {Higgins, D.}, {Koff, R. A.}, {Oey, J.}, {Pilcher, F.}, {Stephens, R.}, {Strabla, L. P.}, {Ulisse, Q.}, \& {Girelli, R.}}]{Hanus2013}
Hanu\v{s}, J., {Bro\v{z}, M.}, {\v{D}urech, J.}, {et~al.} 2013, A\&A, 559, A134, \dodoi{10.1051/0004-6361/201321993}

\bibitem[{{Hanu{\v{s}}} {et~al.}(2013{\natexlab{a}}){Hanu{\v{s}}}, {Bro{\v{z}}}, {Durech}, {Warner}, {Brinsfield}, {Durkee}, {Higgins}, {Koff}, {Oey}, {Pilcher}, {Stephens}, {Strabla}, {Ulisse}, \& {Girelli}}]{2013AA...559A.134H}
{Hanu{\v{s}}}, J., {Bro{\v{z}}}, M., {Durech}, J., {et~al.} 2013{\natexlab{a}}, Astronomy and Astrophysics, 559, A134, \dodoi{10.1051/0004-6361/201321993}

\bibitem[{{Hanu{\v{s}}} {et~al.}(2013{\natexlab{b}}){Hanu{\v{s}}}, {{\v{D}}urech}, {Bro{\v{z}}}, {Marciniak}, {Warner}, {Pilcher}, {Stephens}, {Behrend}, {Carry}, {{\v{C}}apek}, {Antonini}, {Audejean}, {Augustesen}, {Barbotin}, {Baudouin}, {Bayol}, {Bernasconi}, {Borczyk}, {Bosch}, {Brochard}, {Brunetto}, {Casulli}, {Cazenave}, {Charbonnel}, {Christophe}, {Colas}, {Coloma}, {Conjat}, {Cooney}, {Correira}, {Cotrez}, {Coupier}, {Crippa}, {Cristofanelli}, {Dalmas}, {Danavaro}, {Demeautis}, {Droege}, {Durkee}, {Esseiva}, {Esteban}, {Fagas}, {Farroni}, {Fauvaud}, {Fauvaud}, {Del Freo}, {Garcia}, {Geier}, {Godon}, {Grangeon}, {Hamanowa}, {Hamanowa}, {Heck}, {Hellmich}, {Higgins}, {Hirsch}, {Husarik}, {Itkonen}, {Jade}, {Kami{\'n}ski}, {Kankiewicz}, {Klotz}, {Koff}, {Kryszczy{\'n}ska}, {Kwiatkowski}, {Laffont}, {Leroy}, {Lecacheux}, {Leonie}, {Leyrat}, {Manzini}, {Martin}, {Masi}, {Matter}, {Micha{\l}owski}, {Micha{\l}owski}, {Micha{\l}owski}, {Michelet}, {Michelsen}, {Morelle}, {Mottola}, {Naves}, {Nomen}, {Oey},
  {Og{\l}oza}, {Oksanen}, {Oszkiewicz}, {P{\"a}{\"a}kk{\"o}nen}, {Paiella}, {Pallares}, {Paulo}, {Pavic}, {Payet}, {Poli{\'n}ska}, {Polishook}, {Poncy}, {Revaz}, {Rinner}, {Rocca}, {Roche}, {Romeuf}, {Roy}, {Saguin}, {Salom}, {Sanchez}, {Santacana}, {Santana-Ros}, {Sareyan}, {Sobkowiak}, {Sposetti}, {Starkey}, {Stoss}, {Strajnic}, {Teng}, {Tr{\'e}gon}, {Vagnozzi}, {Velichko}, {Waelchli}, {Wagrez}, \& {W{\"u}cher}}]{2013AA...551A..67H}
{Hanu{\v{s}}}, J., {{\v{D}}urech}, J., {Bro{\v{z}}}, M., {et~al.} 2013{\natexlab{b}}, Astronomy and Astrophysics, 551, A67, \dodoi{10.1051/0004-6361/201220701}

\bibitem[{{Hanu{\v{s}}} {et~al.}(2016){Hanu{\v{s}}}, {{\v{D}}urech}, {Oszkiewicz}, {Behrend}, {Carry}, {Delbo}, {Adam}, {Afonina}, {Anquetin}, {Antonini}, {Arnold}, {Audejean}, {Aurard}, {Bachschmidt}, {Baduel}, {Barbotin}, {Barroy}, {Baudouin}, {Berard}, {Berger}, {Bernasconi}, {Bosch}, {Bouley}, {Bozhinova}, {Brinsfield}, {Brunetto}, {Canaud}, {Caron}, {Carrier}, {Casalnuovo}, {Casulli}, {Cerda}, {Chalamet}, {Charbonnel}, {Chinaglia}, {Cikota}, {Colas}, {Coliac}, {Collet}, {Coloma}, {Conjat}, {Conseil}, {Costa}, {Crippa}, {Cristofanelli}, {Damerdji}, {Deback{\`e}re}, {Decock}, {D{\'e}hais}, {D{\'e}l{\'e}age}, {Delmelle}, {Demeautis}, {Dr{\'o}{\.z}d{\.z}}, {Dubos}, {Dulcamara}, {Dumont}, {Durkee}, {Dymock}, {Escalante del Valle}, {Esseiva}, {Esseiva}, {Esteban}, {Fauchez}, {Fauerbach}, {Fauvaud}, {Fauvaud}, {Forn{\'e}}, {Fournel}, {Fradet}, {Garlitz}, {Gerteis}, {Gillier}, {Gillon}, {Giraud}, {Godard}, {Goncalves}, {Hamanowa}, {Hamanowa}, {Hay}, {Hellmich}, {Heterier}, {Higgins}, {Hirsch}, {Hodosan}, {Hren},
  {Hygate}, {Innocent}, {Jacquinot}, {Jawahar}, {Jehin}, {Jerosimic}, {Klotz}, {Koff}, {Korlevic}, {Kosturkiewicz}, {Krafft}, {Krugly}, {Kugel}, {Labrevoir}, {Lecacheux}, {Lehk{\'y}}, {Leroy}, {Lesquerbault}, {Lopez-Gonzales}, {Lutz}, {Mallecot}, {Manfroid}, {Manzini}, {Marciniak}, {Martin}, {Modave}, {Montaigut}, {Montier}, {Morelle}, {Morton}, {Mottola}, {Naves}, {Nomen}, {Oey}, {Og{\l}oza}, {Paiella}, {Pallares}, {Peyrot}, {Pilcher}, {Pirenne}, {Piron}, {Poli{\'n}ska}, {Polotto}, {Poncy}, {Previt}, {Reignier}, {Renauld}, {Ricci}, {Richard}, {Rinner}, {Risoldi}, {Robilliard}, {Romeuf}, {Rousseau}, {Roy}, {Ruthroff}, {Salom}, {Salvador}, {Sanchez}, {Santana-Ros}, {Scholz}, {S{\'e}n{\'e}}, {Skiff}, {Sobkowiak}, {Sogorb}, {Sold{\'a}n}, {Spiridakis}, {Splanska}, {Sposetti}, {Starkey}, {Stephens}, {Stiepen}, {Stoss}, {Strajnic}, {Teng}, {Tumolo}, {Vagnozzi}, {Vanoutryve}, {Vugnon}, {Warner}, {Waucomont}, {Wertz}, {Winiarski}, \& {Wolf}}]{Hanus2016}
{Hanu{\v{s}}}, J., {{\v{D}}urech}, J., {Oszkiewicz}, D.~A., {et~al.} 2016, \aap, 586, A108, \dodoi{10.1051/0004-6361/201527441}

\bibitem[{{Higley} {et~al.}(2008){Higley}, {Hardersen}, \& {Dyvig}}]{2008MPBu...35...63H}
{Higley}, S., {Hardersen}, P., \& {Dyvig}, R. 2008, Minor Planet Bulletin, 35, 63

\bibitem[{{Hung} {et~al.}(2022){Hung}, {Hanu{\v{s}}}, {Masiero}, \& {Tholen}}]{2022PSJ.....3...56H}
{Hung}, D., {Hanu{\v{s}}}, J., {Masiero}, J.~R., \& {Tholen}, D.~J. 2022, The Planetary Science Journal, 3, 56, \dodoi{10.3847/PSJ/ac4d1f}

\bibitem[{Husárik(2016)}]{2016EMP..119...35H}
Husárik, M. 2016, Earth, Moon, and Planets, 119, 35, \dodoi{10.1007/s11038-016-9498-x}

\bibitem[{Ivezic {et~al.}(2020)Ivezic, Juric, Lupton, Tabachnik, Quinn, \& the SDSS~Collaboration}]{ivezic2020sdss}
Ivezic, Z., Juric, M., Lupton, R., {et~al.} 2020, SDSS Moving Object Catalog V1.0, NASA Planetary Data System, \dodoi{10.26033/bv8r-xe89}

\bibitem[{Kryszczy{\'n}ska(2013)}]{Kryszczynska2013}
Kryszczy{\'n}ska, A. 2013, A\&A, 551, A102, \dodoi{10.1051/0004-6361/201220490}

\bibitem[{{Loera-Gonz{\'a}lez} {et~al.}(2023){Loera-Gonz{\'a}lez}, {Olgu{\'\i}n}, {Saucedo}, {Contreras}, {Nu{\~n}ez-L{\'o}pez}, {Dom{\'\i}nguez-Gonz{\'a}lez}, \& {Cortez}}]{Gonz2023MPBu...50..258L}
{Loera-Gonz{\'a}lez}, P.~A., {Olgu{\'\i}n}, L., {Saucedo}, J.~C., {et~al.} 2023, Minor Planet Bulletin, 50, 258

\bibitem[{Marsset {et~al.}(2022)Marsset, DeMeo, Burt, Polishook, Binzel, Granvik, Vernazza, Carry, Bus, Slivan, Thomas, Moskovitz, \& Rivkin}]{Marsset2022}
Marsset, M., DeMeo, F.~E., Burt, B., {et~al.} 2022, The Astronomical Journal, 163, 165, \dodoi{10.3847/1538-3881/ac532f}

\bibitem[{Martikainen {et~al.}(2021)Martikainen, {Muinonen, K.}, {Penttilä, A.}, {Cellino, A.}, \& {Wang, X.-B.}}]{Martikainen2021}
Martikainen, J., {Muinonen, K.}, {Penttilä, A.}, {Cellino, A.}, \& {Wang, X.-B.} 2021, A\&A, 649, A98, \dodoi{10.1051/0004-6361/202039796}

\bibitem[{{Masiero} {et~al.}(2015){Masiero}, {DeMeo}, {Kasuga}, \& {Parker}}]{Masiero2015}
{Masiero}, J.~R., {DeMeo}, F.~E., {Kasuga}, T., \& {Parker}, A.~H. 2015, in Asteroids IV, ed. P.~Michel, F.~DeMeo, \& W.~Bottke (Tucson: University of Arizona Press), 323--340, \dodoi{10.2458/azu_uapress_9780816532131-ch017}

\bibitem[{Milani {et~al.}(2017)Milani, Knežević, Spoto, Cellino, Novaković, \& Tsirvoulis}]{Milani2017}
Milani, A., Knežević, Z., Spoto, F., {et~al.} 2017, Icarus, 288, 240, \dodoi{https://doi.org/10.1016/j.icarus.2016.12.030}

\bibitem[{Muinonen {et~al.}(2020)Muinonen, {Torppa, J.}, {Wang, X.-B.}, {Cellino, A.}, \& {Penttilä, A.}}]{Muinonen2020}
Muinonen, K., {Torppa, J.}, {Wang, X.-B.}, {Cellino, A.}, \& {Penttilä, A.} 2020, A\&A, 642, A138, \dodoi{10.1051/0004-6361/202038036}

\bibitem[{Muinonen {et~al.}(2022)Muinonen, {Uvarova}, {Martikainen}, {Penttil{\"a}}, {Cellino}, \& {Wang}}]{Muinonen2022}
Muinonen, K., {Uvarova}, E., {Martikainen}, J., {et~al.} 2022, Front. Astron. Space Sci., 9, 1, \dodoi{10.3389/fspas.2022.821125}

\bibitem[{Nesvorný {et~al.}(2003)Nesvorný, Bottke, Levison, \& Dones}]{Nesvorny2003}
Nesvorný, D., Bottke, W.~F., Levison, H.~F., \& Dones, L. 2003, The Astrophysical Journal, 591, 486, \dodoi{10.1086/374807}

\bibitem[{Nolan {et~al.}(2013)Nolan, Magri, Howell, Benner, Giorgini, Hergenrother, Hudson, Lauretta, Margot, Ostro, \& Scheeres}]{NOLAN2013629}
Nolan, M.~C., Magri, C., Howell, E.~S., {et~al.} 2013, Icarus, 226, 629, \dodoi{https://doi.org/10.1016/j.icarus.2013.05.028}

\bibitem[{Noonan {et~al.}(2019)Noonan, Reddy, Harris, Bottke, Sanchez, Furfaro, Brown, Fernandes, Kareta, Lejoly, Nallapu, Niazi, Slick, Schatz, Sharkey, Springmann, Angle, Bailey, Acuna, Lewin, Marchese, Meshel, Quintero, Tatum, \& Wilburn}]{Noonan2019}
Noonan, J.~W., Reddy, V., Harris, W.~M., {et~al.} 2019, The Astronomical Journal, 158, 213, \dodoi{10.3847/1538-3881/ab4813}

\bibitem[{Novakovi{\'c} {et~al.}(2017)Novakovi{\'c}, Tsirvoulis, Granvik, \& Todovi{\'c}}]{Novakovic2017}
Novakovi{\'c}, B., Tsirvoulis, G., Granvik, M., \& Todovi{\'c}, A. 2017, The Astronomical Journal, 153, 266, \dodoi{10.3847/1538-3881/aa6ea8}

\bibitem[{{Novakovi{\'c}} {et~al.}(2022){Novakovi{\'c}}, {Vokrouhlick{\'y}}, {Spoto}, \& {Nesvorn{\'y}}}]{Novakovic2022}
{Novakovi{\'c}}, B., {Vokrouhlick{\'y}}, D., {Spoto}, F., \& {Nesvorn{\'y}}, D. 2022, Celestial Mechanics and Dynamical Astronomy, 134, 34, \dodoi{10.1007/s10569-022-10091-7}

\bibitem[{P{\'a}l {et~al.}(2020)P{\'a}l, Szak{\'a}ts, Kiss, B{\'o}di, Bogn{\'a}r, Kalup, Kiss, Marton, Moln{\'a}r, Plachy, S{\'a}rneczky, Szab{\'o}, \& Szab{\'o}}]{pal2020}
P{\'a}l, A., Szak{\'a}ts, R., Kiss, C., {et~al.} 2020, The Astrophysical Journal Supplement Series, 247, 26, \dodoi{10.3847/1538-4365/ab64f0}

\bibitem[{Paolicchi \& Knežević(2016)}]{PAOLICCHI2016314}
Paolicchi, P., \& Knežević, Z. 2016, Icarus, 274, 314, \dodoi{https://doi.org/10.1016/j.icarus.2016.03.005}

\bibitem[{{Pentik{\"a}inen} {et~al.}(2024){Pentik{\"a}inen}, {MacLennan}, {Uvarova}, {Muinonen}, {Penttil{\"a}}, {Wilawer}, {Oszkiewicz}, {Cellino}, {Tanga}, {Wang}, \& {Virkki}}]{pentikainen2024asteroid}
{Pentik{\"a}inen}, H., {MacLennan}, E., {Uvarova}, E., {et~al.} 2024, in European Planetary Science Congress, EPSC2024--894, \dodoi{10.5194/epsc2024-894}

\bibitem[{{Popescu} {et~al.}(2018){Popescu}, {Licandro, J.}, {Carvano, J. M.}, {Stoicescu, R.}, {de Le{\'o}n, J.}, {Morate, D.}, {Boacă, I. L.}, \& {Cristescu, C. P.}}]{Popescu2018}
{Popescu}, M., {Licandro, J.}, {Carvano, J. M.}, {et~al.} 2018, A\&A, 617, A12, \dodoi{10.1051/0004-6361/201833023}

\bibitem[{Riello {et~al.}(2021)Riello, {De Angeli, F.}, {Evans, D. W.}, {Montegriffo, P.}, {Carrasco, J. M.}, {Busso, G.}, {Palaversa, L.}, {Burgess, P. W.}, {Diener, C.}, {Davidson, M.}, {Rowell, N.}, {Fabricius, C.}, {Jordi, C.}, {Bellazzini, M.}, {Pancino, E.}, {Harrison, D. L.}, {Cacciari, C.}, {van Leeuwen, F.}, {Hambly, N. C.}, {Hodgkin, S. T.}, {Osborne, P. J.}, {Altavilla, G.}, {Barstow, M. A.}, {Brown, A. G. A.}, {Castellani, M.}, {Cowell, S.}, {De Luise, F.}, {Gilmore, G.}, {Giuffrida, G.}, {Hidalgo, S.}, {Holland, G.}, {Marinoni, S.}, {Pagani, C.}, {Piersimoni, A. M.}, {Pulone, L.}, {Ragaini, S.}, {Rainer, M.}, {Richards, P. J.}, {Sanna, N.}, {Walton, N. A.}, {Weiler, M.}, \& {Yoldas, A.}}]{GaiaEDR3_2021}
Riello, M., {De Angeli, F.}, {Evans, D. W.}, {et~al.} 2021, A\&A, 649, A3, \dodoi{10.1051/0004-6361/202039587}

\bibitem[{{Sergeyev} {et~al.}(2022){Sergeyev}, {Carry}, {Onken}, {Devillepoix}, {Wolf}, \& {Chang}}]{Sergeyev2022}
{Sergeyev}, A.~V., {Carry}, B., {Onken}, C.~A., {et~al.} 2022, \aap, 658, A109, \dodoi{10.1051/0004-6361/202142074}

\bibitem[{{Skiff} {et~al.}(2023){Skiff}, {McLelland}, {Sanborn}, \& {Koehn}}]{Brian2023MPBu...50...74S}
{Skiff}, B.~A., {McLelland}, K.~P., {Sanborn}, Jason, J., \& {Koehn}, B.~W. 2023, Minor Planet Bulletin, 50, 74

\bibitem[{{Slivan}(2002)}]{Slivan2002}
{Slivan}, S.~M. 2002, \nat, 419, 49, \dodoi{10.1038/nature00993}

\bibitem[{Spoto {et~al.}(2015)Spoto, Milani, \& Knežević}]{SPOTO2015}
Spoto, F., Milani, A., \& Knežević, Z. 2015, Icarus, 257, 275, \dodoi{https://doi.org/10.1016/j.icarus.2015.04.041}

\bibitem[{Stassun {et~al.}(2018)Stassun, Oelkers, Pepper, Paegert, Lee, Torres, Latham, Charpinet, Dressing, Huber, Kane, Lépine, Mann, Muirhead, Rojas-Ayala, Silvotti, Fleming, Levine, \& Plavchan}]{Stassun_2018}
Stassun, K.~G., Oelkers, R.~J., Pepper, J., {et~al.} 2018, The Astronomical Journal, 156, 102, \dodoi{10.3847/1538-3881/aad050}

\bibitem[{{Stephens} {et~al.}(2021){Stephens}, {Coley}, \& {Warner}}]{2021MPBu...48..246S}
{Stephens}, R.~D., {Coley}, D.~R., \& {Warner}, B.~D. 2021, Minor Planet Bulletin, 48, 246

\bibitem[{{Stephens} \& {Warner}(2021)}]{Stephens2021MPBu...48...56S}
{Stephens}, R.~D., \& {Warner}, B.~D. 2021, Minor Planet Bulletin, 48, 56

\bibitem[{Tanga {et~al.}(2023)Tanga, {Pauwels, T.}, {Mignard, F.}, {Muinonen, K.}, {Cellino, A.}, {David, P.}, {Hestroffer, D.}, {Spoto, F.}, {Berthier, J.}, {Guiraud, J.}, {Roux, W.}, {Carry, B.}, {Delbo, M.}, {Dell’Oro, A.}, {Fouron, C.}, {Galluccio, L.}, {Jonckheere, A.}, {Klioner, S. A.}, {Lefustec, Y.}, {Liberato, L.}, {Ordénovic, C.}, {Oreshina-Slezak, I.}, {Penttilä, A.}, {Pailler, F.}, {Panem, Ch.}, {Petit, J.-M.}, {Portell, J.}, {Poujoulet, E.}, {Thuillot, W.}, {Van Hemelryck, E.}, {Burlacu, A.}, {Lasne, Y.}, \& {Managau, S.}}]{Tanga2023}
Tanga, P., {Pauwels, T.}, {Mignard, F.}, {et~al.} 2023, A\&A, 674, A12, \dodoi{10.1051/0004-6361/202243796}

\bibitem[{{Tholen}(1984)}]{Tholen1984PhDT.........3T}
{Tholen}, D.~J. 1984, PhD thesis, University of Arizona

\bibitem[{{{\v{D}}urech} \& {Hanu{\v{s}}}(2023)}]{Durech2023}
{{\v{D}}urech}, J., \& {Hanu{\v{s}}}, J. 2023, \aap, 675, A24, \dodoi{10.1051/0004-6361/202345889}

\bibitem[{{{\v{D}}urech} {et~al.}(2018){{\v{D}}urech}, {Hanu{\v{s}}}, \& {Al{\'\i}-Lagoa}}]{Durech2018AA...617A..57D}
{{\v{D}}urech}, J., {Hanu{\v{s}}}, J., \& {Al{\'\i}-Lagoa}, V. 2018, Astronomy and Astrophysics, 617, A57, \dodoi{10.1051/0004-6361/201833437}

\bibitem[{{{\v{D}}urech} {et~al.}(2016){{\v{D}}urech}, {Hanu{\v{s}}}, {Oszkiewicz}, \& {Van{\v{c}}o}}]{Durech2016AA...587A..48D}
{{\v{D}}urech}, J., {Hanu{\v{s}}}, J., {Oszkiewicz}, D., \& {Van{\v{c}}o}, R. 2016, Astronomy and Astrophysics, 587, A48, \dodoi{10.1051/0004-6361/201527573}

\bibitem[{{{\v{D}}urech} {et~al.}(2019){{\v{D}}urech}, {Hanu{\v{s}}}, \& {Van{\v{c}}o}}]{2019AA...631A...2D}
{{\v{D}}urech}, J., {Hanu{\v{s}}}, J., \& {Van{\v{c}}o}, R. 2019, \aap, 631, A2, \dodoi{10.1051/0004-6361/201936341}

\bibitem[{{{\v{D}}urech} {et~al.}(2020){{\v{D}}urech}, {Tonry}, {Erasmus}, {Denneau}, {Heinze}, {Flewelling}, \& {Van{\v{c}}o}}]{Durech2020AA...643A..59D}
{{\v{D}}urech}, J., {Tonry}, J., {Erasmus}, N., {et~al.} 2020, \aap, 643, A59, \dodoi{10.1051/0004-6361/202037729}

\bibitem[{{Vokrouhlick{\'y}} {et~al.}(2015){Vokrouhlick{\'y}}, {Bottke}, {Chesley}, {Scheeres}, \& {Statler}}]{Vokrouhlicky2015}
{Vokrouhlick{\'y}}, D., {Bottke}, W.~F., {Chesley}, S.~R., {Scheeres}, D.~J., \& {Statler}, T.~S. 2015, in Asteroids IV, ed. P.~Michel, F.~DeMeo, \& W.~Bottke (Tucson: University of Arizona Press), 509--531, \dodoi{10.2458/azu_uapress_9780816532131-ch027}

\bibitem[{Vokrouhlický {et~al.}(2006)Vokrouhlický, Brož, Bottke, Nesvorný, \& Morbidelli}]{VOKROUHLICKY2006}
Vokrouhlický, D., Brož, M., Bottke, W., Nesvorný, D., \& Morbidelli, A. 2006, Icarus, 182, 118, \dodoi{https://doi.org/10.1016/j.icarus.2005.12.010}

\bibitem[{{Warner}(2010)}]{Warner2009}
{Warner}, B.~D. 2010, Minor Planet Bulletin, 37, 112

\bibitem[{Warner {et~al.}(2009)Warner, Harris, \& Pravec}]{WARNER2009134}
Warner, B.~D., Harris, A.~W., \& Pravec, P. 2009, Icarus, 202, 134, \dodoi{https://doi.org/10.1016/j.icarus.2009.02.003}

\bibitem[{Xu {et~al.}(2023)Xu, Wang, Muinonen, Penttilä, Luo, Gu, Sun, Xu, Liu, Xiang, Cao, \& Wang}]{xu2023}
Xu, X., Wang, X., Muinonen, K., {et~al.} 2023, Monthly Notices of the Royal Astronomical Society, 521, 3925, \dodoi{10.1093/mnras/stad765}

\bibitem[{Zhan(2021)}]{zhan2021}
Zhan, H. 2021, Chinese Science Bulletin, 66, 1290, \dodoi{https://doi.org/10.1360/TB-2021-0016}

\end{thebibliography}
\bibliographystyle{aasjournal}

\end{document}